\documentclass[aps,prb,twocolumn,showpacs,amsmath,amssymb]{revtex4-1}

\usepackage{txfonts}
\usepackage{graphicx}
\usepackage{cases}
\usepackage{dcolumn}
\usepackage{bm}
\usepackage{multirow}
\usepackage{float}
\usepackage{tabularx}
\usepackage{color}

\def\eV{\,\textrm{eV}}

 \def\Ry{\,\textrm{Ry}}
\begin{document}

\title{
Electron transport in graphene/graphene side-contact junction by plane-wave multiple scattering method
}  
\author{Xiang-Guo Li$^{1}$, Iek-Heng Chu$^{1}$, X.-G. Zhang$^{1,2}$ and Hai-Ping Cheng$^{1}$}\email{Email: cheng@qtp.ufl.edu; Tel: 352-392-6256}

\affiliation{
$^1$ Department of Physics and Quantum Theory Project, University of Florida, Gainesville, FL 32611\\
$^2$ Center for Nanophase Materials Sciences and Computer Science and Mathematics Division, Oak Ridge National Laboratory, Oak Ridge, TN 37831
}

\begin{abstract}  
Electron transport in graphene is along the sheet but junction devices are often made by stacking different sheets together in a ``side-contact'' geometry which causes the current to flow perpendicular to the sheets within the device. Such geometry presents a challenge to first-principles transport methods. We solve this problem by implementing a plane-wave based multiple scattering theory for electron transport. This implementation improves the computational efficiency over the existing plane-wave transport code, scales better for parallelization over large number of nodes, and does not require the current direction to be along a lattice axis. As a first application, we calculate the tunneling current through a side-contact graphene junction formed by two separate graphene sheets with the edges overlapping each other. We find that transport properties of this junction depend strongly on the AA or AB stacking within the overlapping region as well as the vacuum gap between two graphene sheets. Such transport behaviors are explained in terms of carbon orbital orientation, hybridization, and delocalization as the geometry is varied.

\end{abstract}

\maketitle

\section{\label{sec:1}INTRODUCTION}

Graphene has some of the most fascinating electrical, thermal and mechanical properties \cite{gra1, gra2} that promise to make it an important material for broad applications. However, to realize such a promise we need to learn more than its bulk properties. For example, most large-area graphene films are produced as polycrystalline sheets \cite{grain1, grain2, grain3, grain4} containing multiple small domains, usually connected by one of two types of boundaries, end contacts where two domains connect within the same sheet with direct atomic bonding and usually referred to as grain boundaries \cite{study1,study2,study3,study4,study5},  and side contacts formed by stacking edge regions of the two graphene domains side by side with van der Waals force holding them together, which are also observed in recent experiment \cite{overlap}. There is a great potential for useful devices\cite{junction1,junction2} using side-contact junctions, in which the overlapping region is the device region while the rest of the two graphene domains act as electrodes. First-principles transport study of either types of boundaries in graphene sheets can be challenging because of poor screening due to low dimensionality.  Moreover, the side-contact junctions present a particularly difficult problem because of its unaccommodating geometry for computational methods designed to deal with layer-structured systems.

There are two basic approaches to apply the mesoscopic theory of Laudauer and Buttiker \cite{laudau,buttiker} to study quantum transport of electrons within the first-principles method. The first approach is to use localized basis sets \cite{local1, local2,local3,local4} which allow the calculation of the Green's function of an electrode-device-electrode assembly and a straightforward transport calculation based on the Green's function method. However, localized basis sets do not work well for tunneling through large vacuum gaps that require a faithful description of vacuum electron wave functions. The second approach is based on the scattering theory of plane waves. It has been shown to be completely equivalent to the nonequilibrium Green's function method in the case of noninteracting electrons\cite{green-scatter}. A rigorous first-principles method \cite{pwcond1,pwcond2} (Choi-Ihm method) based on scattering theory and pseudopotentials is implemented within the PWCOND part of the QUANTUM ESPRESSO package \cite{QM}. This code has become one of the standard tools for quantum transport studies \cite{pwcondap1,pwcondap5,pwcondap3,pwcondap4,pwcondap6}. Within this second approach there is also a somewhat different implementation based on the Korringa-Kohn-Rostoker (KKR) band theory, alternatively called the multiple scattering theory, that is adapted to layer-structured systems and thus named the layer-KKR method\cite{green-scatter, lkkrap2}. This method has been particularly successful in the study of spintronics \cite{lkkrapp1,lkkrapp2,lkkrapp3}. While localized basis is inadequate for the study of tunneling current between two graphene sheets in the side-contact junction, neither the Choi-Ihm method nor the layer-KKR method can be effectively applied to this problem as well, as we will discuss below. This dissatisfaction compels us to search for a third implementation of the plane wave scattering method.

Both the layer-KKR and Choi-Ihm methods use a two-dimensional plane-wave basis and divide the system under study into a stack of sufficiently thin slices along the transport direction. A generalized complex band structure \cite{cmband} or transmission matrices are then computed by stitching these slices together with appropriate boundary conditions. Each method has its own advantages and drawbacks. On the one hand, the layer-KKR method first solves the Green's function of individual atomic layers and then stitches these layers  together using a layer doubling technique based on multiple scattering theory. This approach is more efficient and yields  the scattering matrix for any part of the system of interest thus providing more information about the transport properties of the system.  However, it has a serious drawback. It is implemented within the muffin-tin or the atomic sphere approximations (ASA), requiring that the space be divided into spheres around each atom within which the Kohn-Sham potential is spherically symmetric. On the other hand, the Choi-Ihm method does not require spherical approximations. It uses much thinner slices and stitches the slices together by matching boundary conditions of the wave functions between the slices. While this method is in principle rigorous, it is computationally expensive. Moreover, it only yields transport information for the whole system at the end without providing any scattering matrix of the individual parts of the system.

\begin{figure*}[t]
{\includegraphics[width=1.0\textwidth]{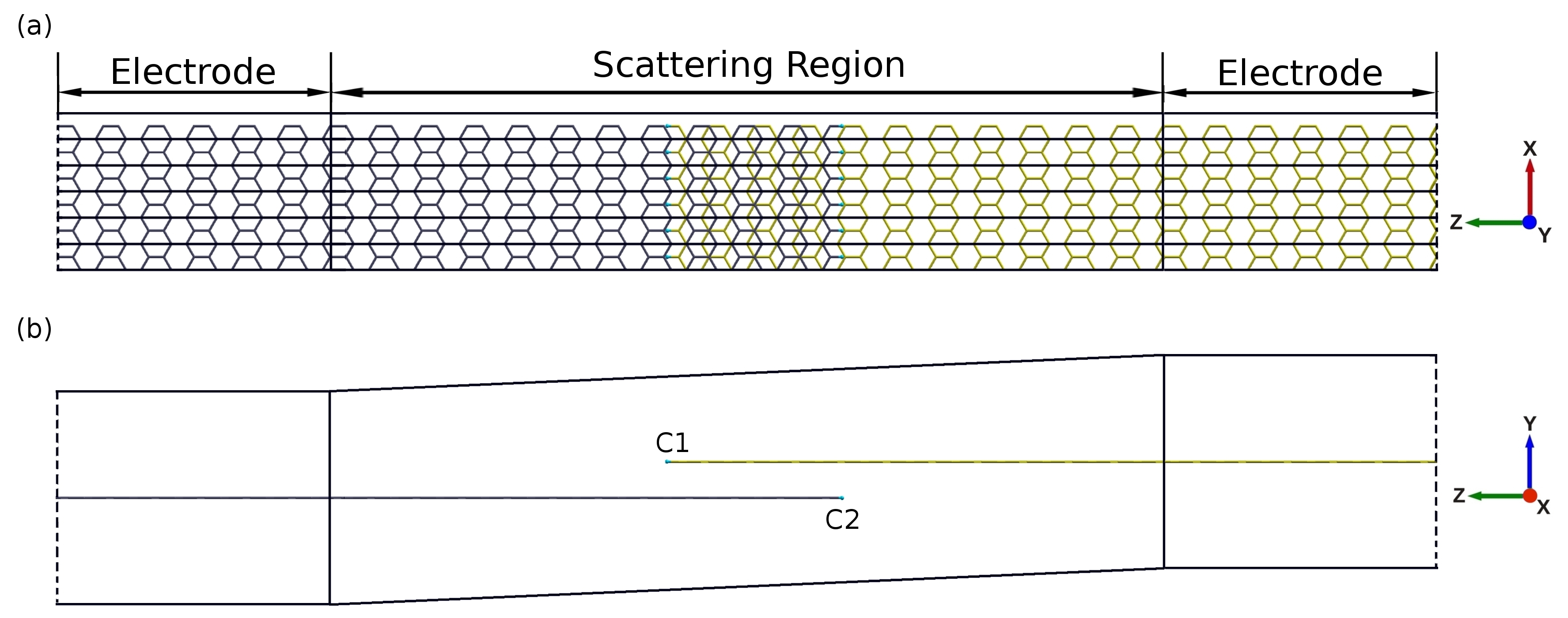}} \caption{\label{fig:str}  
(Color online) Schematic {\bf{(a)}} top and {\bf{(b)}} side views of a graphene side-contact junction. The two graphene layers are connected by one sheet overlapping another to form a bilayer boundary region. The system is periodic along the $x$ direction (six supercells are drawn in the top view) and has a large vacuum gap (more than 15$\AA$) along the $y$ direction. Along the $z$ direction, the system contains the overlapping region and several electrode (graphene) buffer ``layers'' to ensure the convergence of the potential at the boundary to that in bulk electrodes. The unit cell contains one edge carbon atom on each graphene sheet, denoted as C1 (top layer) and C2 (bottom layer), respectively. }
\end{figure*}

Here we present a plane-wave multiple scattering algorithm that combines the advantages of the layer-KKR and Choi-Ihm methods, and show that it can be an efficient first-principles quantum transport method for tunnel junctions represented by the side-contact junction of graphene sheet discussed above and other devices characterized by low symmetry. We first completely reformulate the multiple scattering theory within a plane-wave basis, then implement an algorithm similar to the layer-KKR method but without the muffin-tin or ASA approximations. Additional computational speedup is realized by calculating the complex bands of each electrode using only its two-dimensional primitive unit cell and then folding the complex bands into the smaller Brillouin zone corresponding to the larger two-dimensional supercell of the scattering region (described in Section \ref{sec:2}), and by incorporating the method developed by Srivastava \textit{et al.} \cite{Manoj} for low-symmetry nonorthogonal lattices based on the three-dimensional Bloch theorem. While the impementation of both improvements is straightforward, neither is available in the PWCOND code.  As a first application, we use the new method to study conduction through a graphene/graphene side-contact junction. Our calculations show that there are significant differences in transmission coefficients and their scaling with the overlapping area between AA and AB stacked graphene sheets. This difference is explained in terms of the orbital delocalization and barrier variation for the two geometries.

The paper is organized as follows. An overview of the theory and the computational approach are given in Section {\color {blue}II}. In Section {\color{blue}III}, we discuss the calculation of transport properties of graphene/graphene side-contact junctions and present the results. Conclusions are in Section {\color{blue}IV}. A detailed derivation of the method is provided in Appendices.

\section{\label{sec:2}MODEL AND METHODS}

\subsection{\label{sec:21}Overview of the method }

We consider a system as sketched in FIG. \ref{fig:str}, consisting of a central scattering region connected to left and right semi-infinite electrodes. Mapping this to the graphene side-contact junction, the two electrodes are two semi-infinite graphene sheets with bulk potentials, and the scattering region contains the overlapping region plus a few extra layers outside the overlapping region on both sides to ensure convergence. The plane-wave multiple scattering theory method produces either complex band structure for each of the bulk electrodes or the transmission coefficients for the electrode-device-electrode assembly. The calculation is divided into two stages. In the first stage,
the complex band structures of both electrodes are calculated. Even though the basic formalism for this stage is equivalent between the new method and the Choi-Ihm method, the use of the multiple scattering theory speeds up the new method by about 17\% on a single processor for the same supercell size and allows it to be more efficiently parallelized. A further speedup is achieved by recognizing that the complex
band structures of the electrodes can be computed using primitive cells only, and be folded into a larger supercell that matches the
transverse dimension (perpendicular to the current) of the
scattering region at the end of this stage  (Appendix {\color{blue}B}). In the second stage, a set of linear equations are solved to obtain the transmission coefficients. While the final step of solving the linear system is identical to the Choi-Ihm method, most of the computational time
is spent in the steps needed to set up the final equations. Again implementing these steps using scattering matrices makes the new method significantly more efficient for parallel computation.

To set up the equations, we first apply in each region the two-dimensional periodic boundary conditions in the transverse ($x$, $y$) directions, which are perpendicular to the direction of transport. Systems that are not periodic along the transverse directions can be approximated by sufficiently large two-dimensional supercells with periodic boundary conditions. The Kohn-Sham equation \cite{DFT} within the framework of the plane-wave pseudopotential method is,
\begin{eqnarray}
E\psi({\bf r})\label{kse}&=&-\frac{\hbar^2}{2m}\nabla^2\psi({\bf r})+V_{loc}({\bf r})\psi({\bf r})\nonumber\\
&&+\sum_{\alpha lm}C_{\alpha lm}\sum_{{\bf R}_{\bot}}e^{i{\bf k}_{\bot}\cdot{\bf R}_{\bot}}W_{lm}^{\alpha}({\bf r}-{\boldsymbol {\tau}}^{\alpha}-{\bf R}_{\bot})\label{kse0}\\
C_{\alpha lm}&=&\sum_{uv}D_{lm,uv}\int d^3{\bf r}'[W_{uv}^{\alpha}({\bf r}'-{\boldsymbol \tau}^{\alpha})]^{*}\psi({\bf r}'),\label{calm_0}
\end{eqnarray}
where $V_{loc}({\bf r})$ is the total screened local potential that includes the local part of ionic pseudopotential, electrostatic and exchange-correlation potential due to valence electrons; $W_{lm}^{\alpha}({\bf r})$ are a set of projector functions associated with atom $\alpha$ at position $\boldsymbol{\tau}^{\alpha}$, and form the nonlocal part of the pseudopotential with the coefficients $D_{lm,uv}$ \cite{altrasoft,NCPP}; We use subscript $\bot$ to indicate vectors in the $xy$ plane. ${\bf k}_{\bot}$ and ${\bf R}_{\bot}$ are wave vectors and lattice vectors in the $xy$ plane, respectively.

Following Choi and Ihm \cite{pwcond1}, we divide each scattering or electrode region into slices perpendicular to the transport direction ($z$). The slices are sufficiently thin so that within each slice the local potential can be treated as independent of $z$. 
If one such region starting from $z_0=0$ and ending at $z_{N}=d$ is divided into $N$ slices, then Eq. (\ref{kse0}) can be rewritten for each slice (labeled by superscript $p$) as,  
\begin{eqnarray}
E\psi^p({\bf r})\label{kse}&=&-\frac{\hbar^2}{2m}\nabla^2\psi^p({\bf r})+V_{loc}^p({\bf r}_{\bot})\psi^p({\bf r})\nonumber\\
&&+\sum_{\alpha lm}C_{\alpha lm}\sum_{{\bf R}_{\bot}}e^{i{\bf k}_{\bot}\cdot{\bf R}_{\bot}}W_{lm}^{\alpha}({\bf r}-{\boldsymbol {\tau}}^{\alpha}-{\bf R}_{\bot}).\label{ksep}
\end{eqnarray}
The solutions of the above inhomogeneous differential equation for each slice are to be connected together using the boundary conditions that the wave functions and their derivatives are continuous to yield the total wave function $\psi(\bf{r})$ of the entire region in Eq. (\ref{kse0}).
The first step is to find a set of basis functions,
\begin{equation}
\psi_{n}^p({\bf r})=\phi^{p}_{n}({\bf r}_{\bot})e^{\pm ik^{p}_{n}(z-z_{p})}
\end{equation}
satisfying  a homogeneous equation obtained by setting all $C_{alm}$'s in Eq. (\ref{ksep}) to zero, where $n$ labels different homogeneous solutions and goes up to $N_{2D}$, the cutoff number of the reciprocal lattice vectors ${\bf G}_{\bot}$ in the $xy$ plane, $k^{p}_{n}=\sqrt{2m(E-E^{p}_{n})/\hbar}$ is the $n^{th}$ wave vector along $z$ direction in the $p^{th}$ slice, $E$ is the total incident energy and $E^{p}_{n}$ is the energy eigenvalue of the following equation,
\begin{eqnarray}
 E^{p}_{n}\phi^{p}_{n}({\bf G}_{\bot})&=&\frac{\hbar^2}{2m}|{\bf k}_{\bot}+{\bf G}_{\bot}|^2\phi^{p}_{n}({\bf G}_{\bot})\nonumber \\
&&+\sum_{{\bf G}'_{\bot}}V_{loc}^{p}({\bf G}_{\bot}-{\bf G}'_{\bot})\phi^{p}_{n}({\bf G}'_{\bot})\label{loc},
\end{eqnarray}
where $\phi^{p}_{n}({\bf G}_{\bot})$ is the Fourier transform of $\phi^{p}_{n}({\bf r}_{\bot})$. 
In addition to the homogeneous solution basis set, we also need a particular solution basis set as described below. Starting from Eq. (\ref{ksep}) we set only one of $C_{alm}$'s to one and all others to zero,
\begin{eqnarray}
E\psi_{\alpha lm}^p({\bf r})&=&-\frac{\hbar^2}{2m}\nabla^2\psi_{\alpha lm}^p({\bf r})+V_{loc}^p({\bf r}_\bot)\psi_{\alpha lm}^p({\bf r})\nonumber \\&&+\sum_{{\bf R}_{\bot}}e^{i{\bf k}_{\bot}\cdot{\bf R}_{\bot}}W_{lm}^{\alpha}({\bf r}-{\tau}^{\alpha}-{\bf R}_{\bot}),\label{nloc}
\end{eqnarray}
which yields a particular solution, $\psi_{\alpha lm}^p({\bf r})$, for each $(\alpha, l, m)$. 
Both the homogeneous and the inhomogeneous solution basis sets satisfy the Bloch boundary condition within the $xy$ plane, $\psi_{n\,(\alpha lm)}({\bf r}+{\bf R}_{\bot})=e^{i{\bf k}_{\bot}\cdot{\bf R}_{\bot}}\psi_{n\,(\alpha lm)}({\bf r})$. The general solution of Eq. (\ref{ksep}) is written as a linear combination of $\psi^p_n$ and $\psi^p_{\alpha lm}$,
\begin{eqnarray}
\psi^{p}_{k_i}({\bf r})&=&\sum_{n}A^{p}_{n,k_i}\phi^{p}_{n}({\bf r}_{\bot})e^{ik^{p}_{n}\cdot (z-z_{p})}+\sum_{n}B^{p}_{n,k_i}\phi^{p}_{n}({\bf r}_{\bot})e^{-ik^{p}_{n}\cdot (z-z_{p})}\nonumber\\
&&+\sum_{\alpha lm}C_{\alpha lm,k_i}\psi^{p}_{\alpha lm}({\bf r}),\label{gsol_p0}
\end{eqnarray}
where $k_i$ labels the wavevector of the Bloch and evanescent states when computing complex band structure, and labels the wavevector of the incident waves of the entire scattering region when computing the transmission coefficients. 

The coefficients $A^{p}_{n,k_i}$ and $B^{p}_{n,k_i}$, which depend on $k_i$,  are the only unknowns in the wave function
and are determined by matching the wave functions between adjacent slices, usually in a transfer matrix \cite{tranM} formulation. However, a transfer matrix approach is usually numerically unstable because of the
appearance of the exponential factors in Eq. (\ref{gsol_p0}). For $k^{p}_{n}$ containing nonzero imaginary parts (corresponding to evanescent states), these terms are exponentially decaying and growing waves. Therefore, some transfer matrix elements, as given by Eq. (\ref{transfer}), will grow exponentially while some others will decay exponentially, creating a numerically unstable system. As iteration proceeds, information for the decaying modes will be lost, causing the numerical solution to diverge. To avoid this numerical instability, we separate the waves into forward and backward waves. The forward waves are those that propagate or decay in the positive $z$ direction, the backward waves are those that propagate or decay in the negative $z$ direction. By iterating both types of waves along their exponentially decaying directions we can avoid the numerical instability, a practice already adopted in a number of previous studies\cite{num, scat}. This is accomplished by rearranging the boundary conditions between the slices in terms of incident waves of slice $p+1$ with coefficients $A^p$ and $B^{p+1}$, the outgoing waves of slice $p+1$ with coefficients $A^{p+1}$ and $B^{p}$ using the scattering matrix\cite{scat1,scat2} $S(p,p+1)$ [see  Eq. (\ref{scatter})] that couples them,
\begin{eqnarray}
\left[ \begin{array}{c} A^{p+1} \\ B^{p} \end{array} \right] &=&\left[\begin{array}{c c} S_{11}(p,p+1)\ \ S_{12}(p,p+1)\\ S_{21}(p,p+1)\ \ S_{22}(p,p+1)\end{array}\right] \left[ \begin{array}{c} A^{p} \\ B^{p+1} \end{array} \right]\nonumber \\&&+\left[\begin{array}{c} h^{a}(p,p+1)C\\h^{b}(p,p+1)C\end{array} \right].\label{smm2}
\end{eqnarray}
The coefficient $h(p,p+1)$ is also defined in Eq. (\ref{scatter}). Clearly, $S(p,p)=I$, $h^{a}(p,p)=0$ and $h^{b}(p,p)=0$. 

Calculating the scattering matrix for each slice is the first step to obtain the total scattering matrix $S(1,N)$ of the entire region. The second step is to stitch the slices together by applying a doubling technique similar to that employed in the layer-KKR method. In general, one can obtain the scattering matrix for a collection of slices $m$ through $k$ by combining the scattering matrices for $m$ through $n$ and for $n$ through $k$ ($m<n<k$) using the following multiple scattering equations\cite{lkk},
\begin{equation}
\begin{aligned}
S_{11}(m,k)&=S_{11}(n,k)[I-S_{12}(m,n)S_{21}(n,k)]^{-1}S_{11}(m,n),\\
S_{12}(m,k)&=S_{12}(n,k)+S_{11}(n,k)[I-S_{12}(m,n)S_{21}(n,k)]^{-1}\times \\
                     &\quad S_{12}(m,n)S_{22}(n,k),\\
S_{21}(m,k)&=S_{21}(m,n)+S_{22}(m,n)S_{21}(n,k)\times \\ 
&\quad[I-S_{12}(m,n)S_{21}(n,k)]^{-1}S_{11}(m,n),\\
S_{22}(m,k)&=S_{22}(m,n)S_{22}(n,k)+S_{22}(m,n)S_{21}(n,k)\times \\ 
&\quad[I-S_{12}(m,n)S_{21}(n,k)]^{-1}S_{12}(m,n)S_{22}(n,k),\\
h^{a}(m,k)&=h^{a}(n,k)+S_{11}(n,k)[I-S_{12}(m,n)S_{21}(n,k)]^{-1}\times \\ 
&\quad[S_{12}(m,n)h^{b}(n,k)+h^{a}(m,n)],\\
h^{b}(m,k)&=h^{b}(m,n)+S_{22}(m,n)h^{b}(n,k)+S_{22}(m,n)\times \\
&\quad S_{21}(n,k)[I-S_{12}(m,n)S_{21}(n,k)]^{-1}\times \\
&\quad [S_{12}(m,n)h^{b}(n,k)+h^{a}(m,n)].\label{recurrence1}
\end{aligned}
\end{equation}
Each time these equations are applied, the number of slices represented by the scattering matrix is doubled, thus the name ``layer-doubling''. The final scattering matrix $S(1,N)$ is obtained by applying layer-doubling repeatedly until all slices are contained within the scattering matrix. 

While the scattering matrix formalism is numerically more stable than the transfer matrix formalism, the latter is computationally faster than the former. To achieve the optimal balance between speed and numerical stability, we iterate two of the transfer matrices $I_{11}$ and $I_{12}$ along with the scattering matrices $S_{21}$ and $S_{22}$ (see Eq. (\ref{recurrence2})) for the first few doubling steps, when the condition number of the transfer matrix $I_{11}$ is reasonable. As soon as the condition number of $I_{II}$ turns bad, we switch over to iterate entirely with the scattering matrices  $S_{11}$, $S_{12}$, $S_{21}$ and $S_{22}$.

 The final inhomogeneous equation is identical in form to Eq. (\ref{smm2}),
\begin{eqnarray}
&&\left[ \begin{array}{c} A^{N} \\ B^{1} \end{array} \right] =\left[\begin{array}{c c} S_{11}(1,N)\ \ S_{12}(1,N)\\ S_{21}(1,N)\ \ S_{22}(1,N)\end{array}\right] \left[ \begin{array}{c} A^{1} \\ B^{N} \end{array} \right]+ \left[\begin{array}{c} h^{a}(1,N)C\\h^{b}(1,N)C\end{array} \right]. 
\nonumber\\ \label{smm3_0}
\end{eqnarray}
Eq. (\ref{smm3_0}) provides a direct connection between the total incident waves of the entire region with coefficients $A^1$ and $B^N$ and the total outgoing waves of the entire region with coefficients $A^N$ and $B^1$.

There are $4N_{2D}+N_{orb}$  unknown coefficients counting $A$ ($2N_{2D}$), $B$ ($2N_{2D}$) and $C$ ($N_{orb}$), with $N_{orb}$ being the total number of nonlocal spheres characterized by $W^{\alpha}_{lm}$. As mentioned above, the total wave function of the entire region in Eq. (\ref{kse0}) can be obtained by connecting the wave function in each slice [see Eq. (\ref{gsol_p0})], which means that the total wave function can be expressed as a function of the unknown coefficients $A$, $B$ and $C$. Therefore, the definition of $C$ in Eq. (\ref{calm_0}) provides $N_{orb}$ equations about the unknown coefficients. In addition, Eq. (\ref{smm3_0}) gives us another $2N_{2D}$ equations. So a total of $2N_{2D}+N_{orb}$ linear equations are provided by Eqs. (\ref{calm_0}) and (\ref{smm3_0}). Additional $2N_{2D}$ equations are needed, which are provided by the boundary conditions below.
Two different types of boundary conditions are needed depending on the physical problem under study. For complex band structures, the generalized Bloch boundary conditions \cite{bd-zhang} are employed along the transport direction (see Appendix B). For the transmission coefficients of the scattering region, continuity conditions for the wave function and its derivative are applied at the interfaces between the scattering region and each side of the electrodes (see Appendix C).

To compute the complex band structure of an electrode region, the generalized Bloch conditions along the $z$-direction [see Eqs. (\ref{bc1}), (\ref{bc2}) in Appendix  {\color{blue}B}] need to be applied as the boundary conditions. With Eq. (\ref{smm3_0}), we can express $A^N$ and $B^1$ as a function of $A^1$, $B^N$ and $C$. In addition, we can denote the coefficients $C$ with nonlocal spheres completely fitting the electrode region as $C_{(\alpha lm)^\prime}$, and those with nonlocal spheres crossing the boundaries of the electrode region as $C_{(\alpha lm)}$.  From the definition of $C_{(alm)^\prime}$ in Eq. (\ref{calm_0}), we can express $C_{(\alpha lm)^\prime}$ as a function of $A^N$, $B^1$, $A^1$, $B^N$ and $C_{(\alpha lm)}$. Therefore, Eq. (\ref{smm3_0}) and the definition of $C_{(\alpha lm)^\prime}$ in Eq. (\ref{calm_0}) give us an expression of $A^N$, $B^1$, $C_{(\alpha lm)^\prime}$ as a function of $A^1$, $B^N$ and $C_{(\alpha lm)}$. So in the definition of $C_{(\alpha lm)}$ [Eqs. (\ref{calm2}) and (\ref{calm3})] and Bloch condition Eqs. (\ref{bc1}) and (\ref{bc2}), we can substitute $A^N$, $B^1$, $C_{(\alpha lm)^\prime}$ with $A^1$, $B^N$ and $C_{(\alpha lm)}$.  As a result, the only unknowns in Eqs. (\ref{calm2}) and (\ref{calm3}), and  (\ref{bc1}) and (\ref{bc2}) are $A^1$, $B^N$ and $C_{(\alpha lm)}$.  Rearranging these equations by setting the unknowns $A^1$, $B^N$ and $C_{(\alpha lm)}$ as $X$, we can obtain the generalized eigenvalue problem that takes the form
\begin{eqnarray}
P\cdot X=e^{ikd}e^{i{\bf k}_{\bot}\cdot{\bf a_3}_{\bot}}Q\cdot X,\label{cmp}
\end{eqnarray}
where $\bf a_3$ is the third lattice vector which is not in the $xy$ plane; ${\bf k}=({\bf k}_{\bot},k)$ is the wave vector.
Solving this equation yields a set of forward waves $\{\psi^{+}_{k}({\bf r})\}$, backward waves $\{\psi^{-}_{k}({\bf r})\}$ and also complex band structure $k(E)$. 

For transmission coefficient calculations, we need to match the boundary conditions at  $z_0=0$ and $z_{N}=d$ for the wavefunction and its derivative [see Eqs. (\ref{bound})-(\ref{C3}) in Appendix  {\color{blue}C}]. Through this process, $A^1$ and $B^1$ can be expressed as a function of $A^0$, $B^0$ and $C$ [see Eq. (\ref{left})]; $A^N$ and $B^N$ can be expressed as a function of $A^{N+1}$, $B^{N+1}$ and $C$ [see Eq. (\ref{right})], where $A^0$ and $B^0$, $A^{N+1}$ and $B^{N+1}$ are the wave coefficients in the $0^{th}$ and $(N+1)^{th}$ slices, as defined in Appendix {\color{blue}C}. Therefore, the unknowns in the definition of $C$ of Eq. (\ref{calm_0}), and  Eqs. (\ref{bound})-(\ref{C3}) are $A^{N+1}$, $B^0$ and $C$ (noting that $A^0$ and $B^{N+1}$ specify the incident waves, which are known). Rearranging these equations by setting the unknowns $A^{N+1}$, $B^0$ and $C$ as $X$, we can write a set of  linear equations in the matrix form 
\begin{eqnarray}
M\cdot X=D.\label{trans}
\end{eqnarray}
The reflection and transmission matrices can be obtained from the solution of $X$ in Eq. (\ref{trans}). The details about the dimension of matrices $P$, $Q$, $M$, $D$ as well as the elements of $X$ are discussed in Appendices {\color{blue}B} and {\color{blue}C}. 

Next we compare the efficiency of this method against that of the Choi-Ihm method. In the first stage, in which the complex band structures of the electrodes are calculated, there are two speedups. First, by using the primitive cells of the electrodes, the dimension of Eq. (\ref{cmp}) for complex band calculations can be reduced by up to several fold for a large calculation. This strategy can also be implemented in the
original Choi-Ihm method, although it is not yet available in PWCOND. Second, on a single processor, the new method speeds up the complex band calculations by about 17\% according to benchmark runs. There is an additional speedup in a parallel calculation. Since this
parallel speedup is the same for both stages, we describe it specifically for the
second stage. The most time consuming parts for either stages are the steps to set up Eq. (\ref{cmp}) for the complex band calculation or Eq. (\ref{trans}) for the transmission coefficient calculation, not the final step of solving either equation (which is parallelized efficiently but does not consume much computational time). Both the new method and the Choi-Ihm method have identical Eqs. (\ref{loc}) and (\ref{nloc}) used at the beginning of the calculation, for which nearly perfect parallelization can be achieved. The major difference between the two methods is in the steps immediately following this calculation. In the multiple scattering method, the scattering matrix for each slice [Eq. (\ref{smm2})] is computed, then the doubling technique [Eq. (\ref{recurrence1})] is applied to obtain the scattering matrix [Eq. (\ref{smm3_0})] for the entire system. In the Choi-Ihm method, the wave functions in individual slices are matched across the boundaries in a layer-doubling process until the wave function of the entire system is obtained. 
In both methods the bottleneck for parallelization is the layer-doubling step, which can scale at best as $\log_2 L$ where $L$ is the number of processors used. However, while the Choi-Ihm method relies entirely on the layer-doubling technique for the wave functions, in the multiple scattering method much of the computation load is shifted to the extraction of the scattering matrices for individual layers, which can be parallelized with nearly 100\% efficiency. Once the scattering matrices of all slices are calculated, evaluating Eq. (\ref{recurrence1}) is twice faster than matching boundary conditions for the wave functions across layer boundaries. Therefore the prefactor of the $\log_2 L$
term for the new method is half of that for the Choi-Ihm method.

\subsection{\label{sec:22}Models and computational details }

We apply the plane-wave-multiple-scattering transport method to study the transport properties of a graphene/graphene side contact junction, as illustrated in FIG. \ref{fig:str}. The edges of the graphene sheets are zigzag H-terminated in the supercell, the unit cell contains one edge carbon atom for each sheet with a dangling bond saturated by the H atom. We denote these edge atoms as C1 (top graphene layer) and C2 (bottom graphene layer), respectively.  Transmission and reflection coefficients of the incident electrode Bloch states through the scattering region are calculated by applying the scattering boundary conditions, Eqs. (\ref{bound}-\ref{C3}). The electronic structures of the electrodes and the scattering region are calculated separately using the QUANTUM ESPRESSO package to obtain the self-consistent potentials. Fourteen unit cells of graphene outside the overlapping region are added to the scattering region on each side to ensure the convergence of the potential at the boundaries of the scattering region (see FIG. \ref{fig:str}). The electronic structure calculation for the scattering region is obtained by repeating the region in the $z$ direction. To achieve translational invariance under an unit lattice vector along the $z$ direction, the graphene sheet on the left must coincide with the graphene sheet on the right after translation. The best way to accomplish this is to tilt the supercell such that the lattice vector $\bf a_3$ of the scattering region is no longer perpendicular to the $xy$ plane (see FIG. \ref{fig:str}b). Such a configuration cannot be handled by the existing PWCOND code as we discussed earlier. The PWCOND code can only treat this system by doubling the size of the supercell to make the repeating lattice vector perpendicular to the $xy$ plane, thus incurring an 8-fold increase in computation time and 4-fold increase in memory requirement.

We use the Rappe-Rabe-Kaxiras-Joannopoulos ultrasoft pseudopotentials \cite{rrkj} for the C, H atoms with the Perdew-Burke-Ernzerhof exchange-correlation functional \cite{pbe}. Because of the zigzag H-terminated edges which have non-zero magnetization\cite{ZGNR}, the calculation is spin-polarized. The energy cutoffs are 60$\Ry$ and 360$\Ry$ for the wave function and charge density, respectively. Gaussian smearing with a width of 0.05$\eV$ is used for the energy levels. The scattering region is periodic along the $x$ direction with a thick vacuum layer of more than 15$\,\AA$ in the $y$ direction. A 2.46$\,\AA$ $\times$ 20$\,\AA$ rectangular area is used for the supercell's $xy$ plane. For self-consistent calculations, the $z$ dimension of the supercell is much larger because of the added unit cells of electrodes, and also depends on the width of the overlapping region. A $20\times 1 \times 1$ k-space mesh is sufficient to sample the Brillouin zone (BZ) for the scattering region.

Graphene has a zero density of states (DOS) at the Dirac point. Using it as electrodes requires the transport calculations to be performed at an energy away from the Dirac point to ensure a finite transmission. In our calculation the incident electron energy is  chosen to be $0.1$ eV above the Dirac point. At this energy, electrode conduction channels only exist within a small volume of the reciprocal space. The reciprocal space is discretized as $k_\bot=(m^x\Delta k_x,m^y\Delta k_y)$, where $\Delta k_{x(y)}$ is the mesh interval along $x(y)$ direction, and $m^{x(y)}$ runs from $N_i^{x(y)}$ to $N_f^{x(y)}$. The number
of $k_\bot$ points along the $x(y)$ direction in the reciprocal space is $N^{x(y)}=N_f^{x(y)}-N_i^{x(y)}+1$, which defines the number of  conduction channels in the electrode. The transmission coefficient T($m^x\Delta k_x,m^y\Delta k_y$) for each mesh point is computed using the method described in Section \ref{sec:2}. Because the supercell along the $y$ direction is the vacuum layer of $y$ direction in the electrode in our model system is sufficiently large that the total energy is independent of $k_y$,  the transmission coefficients are also independent of $k_y$. 

Because of the small number of conduction channels in graphene, even if the junction does not contain any scattering the total transmission is still small, a situation we refer to as ``electrode-limited'' conduction. In order to distinguish the effect of junction scattering from that of the electrode-limited conduction, we compute a transmission probability per electrode conduction channel as follows:
\begin{eqnarray}
\bar{T}=\frac{1}{N^x N^y}\sum_{m^x=N_{i}^x}^{N_f^x}\sum_{m^y=N_{i}^y}^{N_f^y}T(m^x\Delta k_x,m^y\Delta k_y)
\label{T1_d}.
\end{eqnarray}
Eq. (\ref{T1_d}) can be reduced to the following form since the transmission coefficients are independent of $k_y$:
\begin{eqnarray}
\bar{T}=\frac{1}{N^x}\sum_{m^x=N_{i}^x}^{N_f^x}T(m^x\Delta k_x,0).\label{T2_d}
\end{eqnarray}
We find that $\Delta k_x=4.167\times10^{-4}(2\pi/a_1)$ is sufficient to converge $\bar{T}$ to $1\%$. Here $a_1$ is the lattice constant of electrode in $x$ direction. In particular, if the $k$ mesh in the reciprocal space is infinitesimal or continuous, Eqs. (\ref{T1_d}) and (\ref{T2_d}) can be rewritten in an integral form:
\begin{eqnarray}
\bar{T}=\frac{1}{\Omega_{\rm eff}}\int_{\Omega_{\rm eff}}T(k_x,k_y)dk_x dk_y,\label{T2}\\
\bar{T}=\frac{1}{\lambda_{\rm eff}^x}\int_{\lambda_{\rm eff}^x}T(k_x,0)dk_x.\label{T1}
\end{eqnarray}
where the effective projected Fermi area of the electrode, denoted as $\Omega_{\rm eff}$, is the area projected to $xy$ plane in $k$ space from the effective Fermi volume of the electrode; $\lambda_{\rm eff}^{x}$ is the effective projected Fermi length in $x$ direction of $k$ space for the electrode graphene. 

\begin{figure*}[t]
{\includegraphics[width=0.9\textwidth]{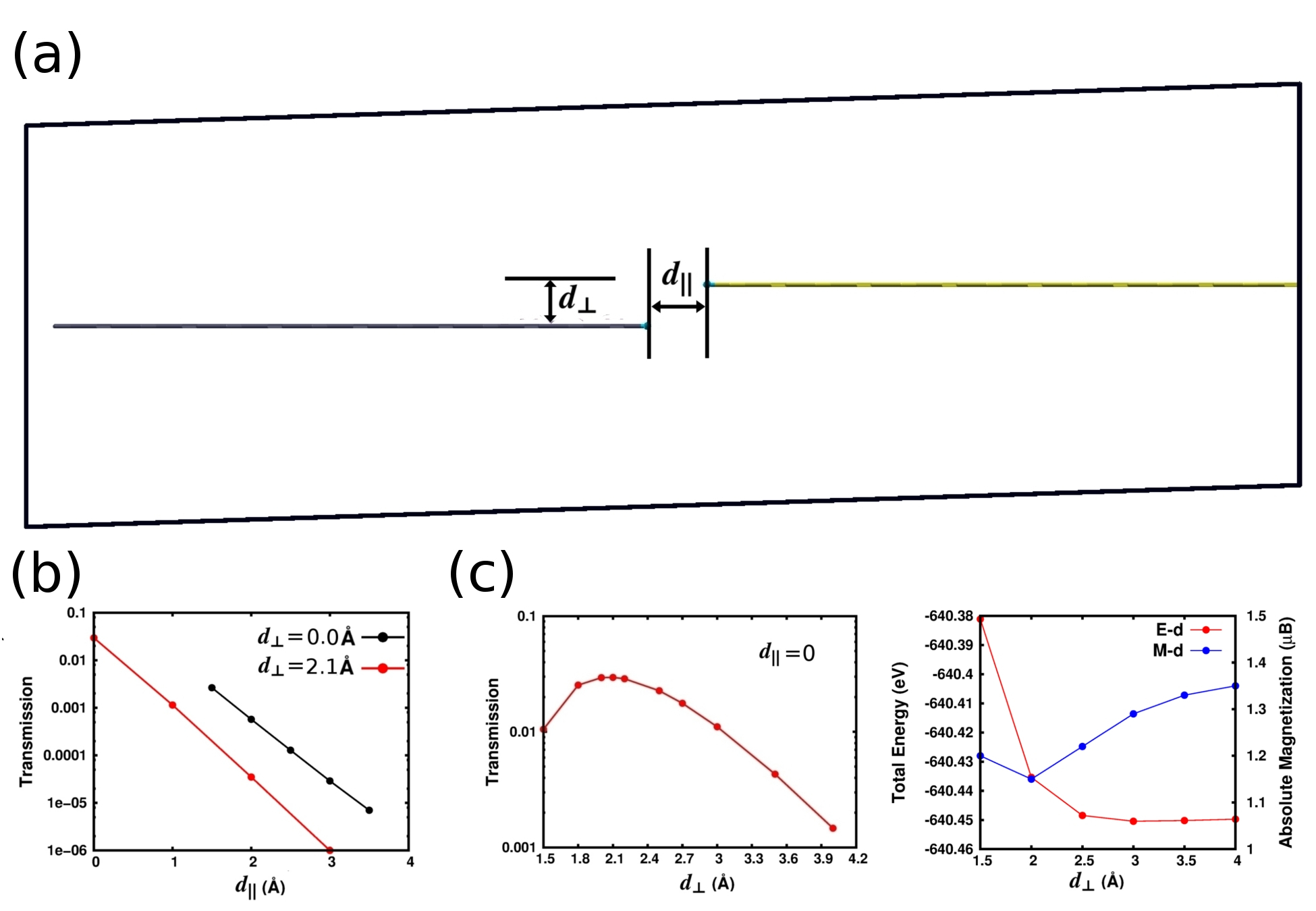}} \caption{\label{fig:dis}  
(Color online) {\bf{(a)}} Schematic side view of two layer graphene with horizontal distance $d_\parallel$ (parallel with graphene plane) and interlayer distance $d_\perp$ (perpendicular to graphene plane). The supercell is tilted so that the left of the bottom layer graphene can match the right of the top graphene layer with a periodic boundary condition. {\bf{(b)}} Transmission as a function of $d_\parallel$ with $d_\perp= 0.0$ \AA\ and $2.1$ \AA . {\bf{(c)}} Transmission for spin-up channel in the left panel and total energy ($E$), absolute magnetization ($M$) in the right panel as a function of $d_\perp$ with $d_\parallel=0$. $y$ axis in both {\bf{b}} and {\bf{c}} is in logarithm scale.}
\end{figure*}

\section{\label{sec:3}Results and Discussion}

\begin{figure*}[t]
{\includegraphics[width=1.0\textwidth]{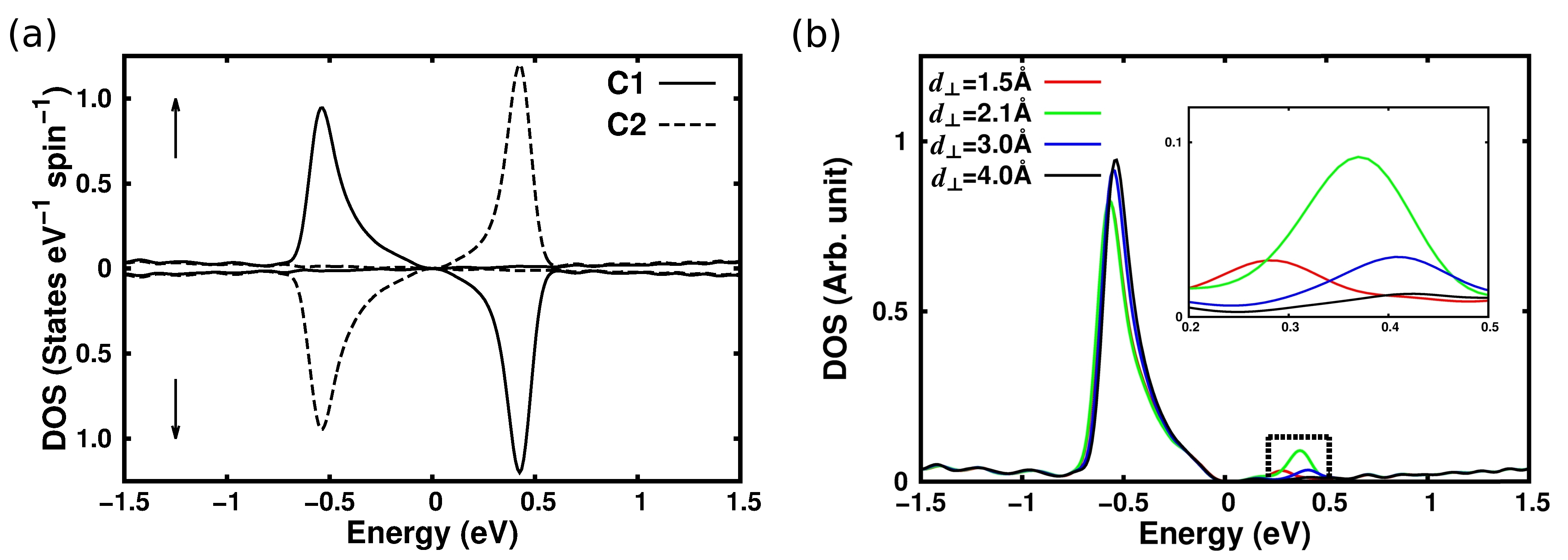}} \caption{\label{fig:pdos}  
(Color online) \textbf{(a)} Projected density of states (PDOS) for the $p$ orbitals of the edge carbon atoms, C1 and C2, for $d_\perp=4.0$\ \AA. At this distance, there is little evidence of coupling between the two graphene sheets. \textbf{(b)} Spin-up PDOS for the $p$ orbital of C1 at different interlayer distances. Here we use a smearing of $\sigma=0.005$\Ry, and the Fermi energy is at $0$ eV.}
\end{figure*}

\begin{figure}[h]
{\includegraphics[width=0.5\textwidth]{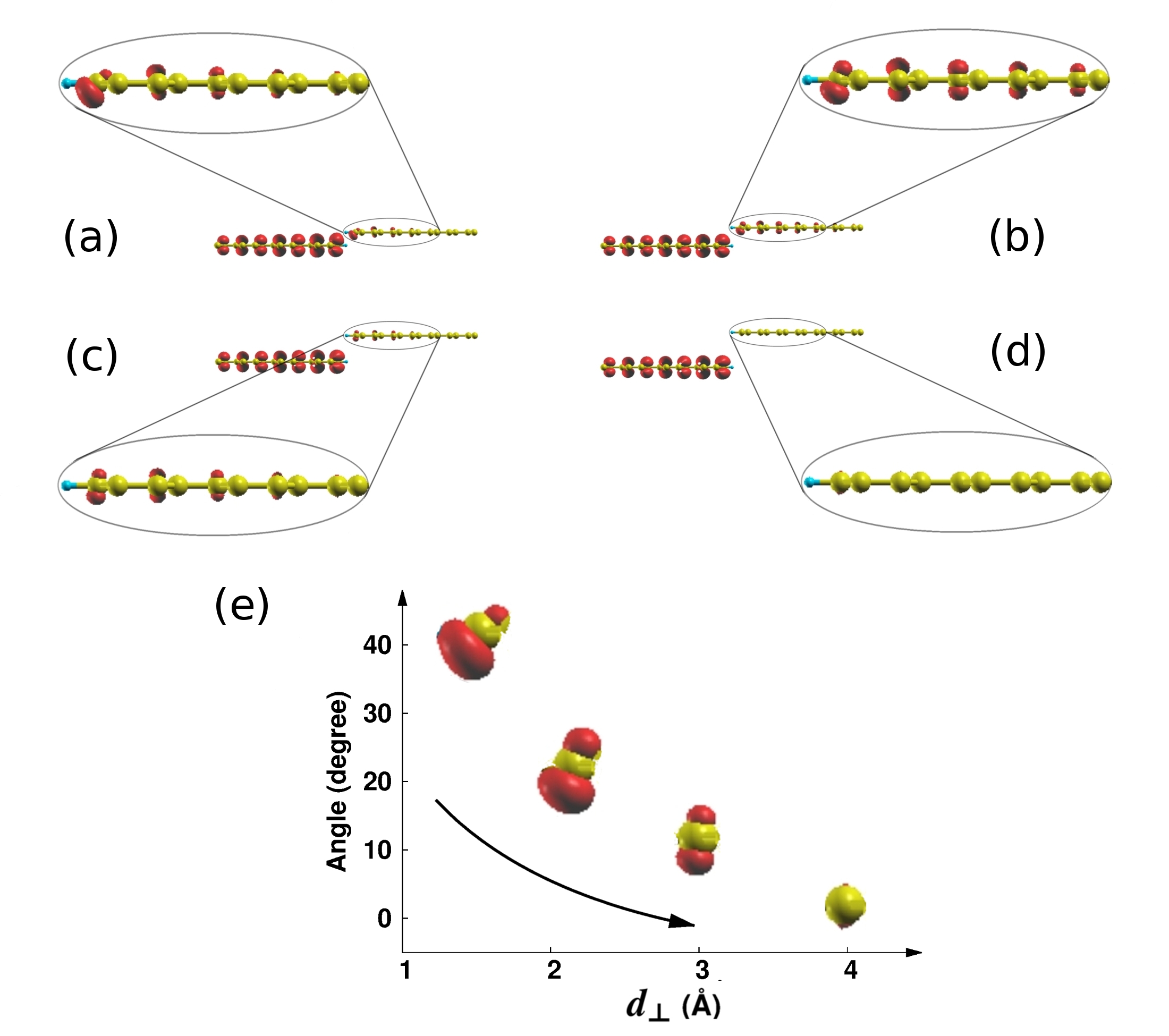}} \caption{\label{fig:ldos}  
(Color online) Isosurfaces (at 0.0002 states/\AA$^3$, in red color) of integrated spin-up local density of states from 0.05$\eV$ below the incident energy to 0.05$\eV$ above the incident energy for two graphene layers with interlayer distance \textbf{(a)} 1.5\,\AA, \textbf{(b)} 2.1\,\AA, \textbf{(c)} 3.0\,\AA, \textbf{(d)} 4.0\,\AA. \textbf{(e)} The tilting angle of the $p_z$ orbital in C1 of the top sheet as a function of the interlayer distance. Yellow balls are carbon atoms and blue balls are hydrogen atoms.}
\end{figure}

\begin{figure*}
{\includegraphics[width=0.9\textwidth]{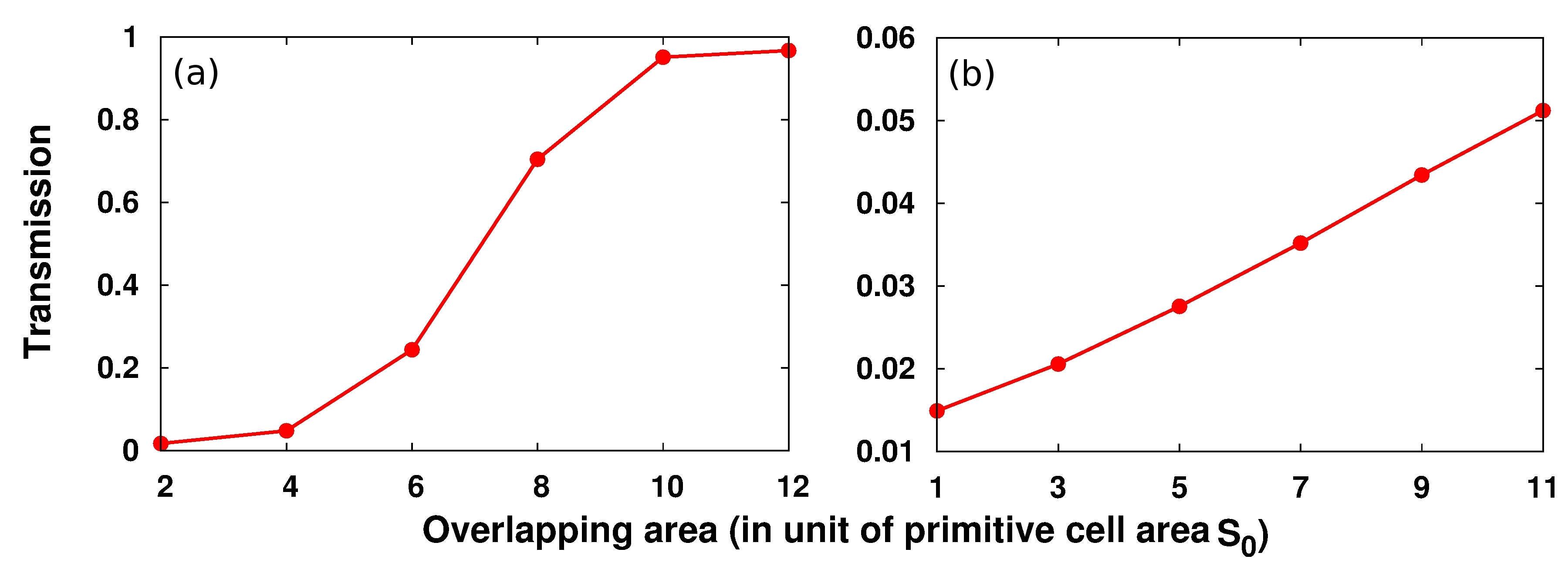}} \caption{\label{fig:overlap}  
(Color online) Transmission as a function of overlapping area for \textbf{(a)} AA stacking and  \textbf{(b)} AB stacking. 
 }
\end{figure*}

A side contact formed between two graphene sheets has already been observed experimentally and is found to be more resistive than grain boundaries formed between graphene domains within the same sheet \cite{study4}. The conduction through the side contact can be greatly improved by increasing the overlapping area in the experimental samples. Therefore, the effect on the transport properties of a side contact junction due to the overlapping area as a function of the interlayer distance between the two graphene sheets is the focus of this first-principles study. We begin with the limiting case that the two graphene domains have no overlapping with each other. In this case the edges play an important role,
where the dominant factor is the orbital hybridization between the states from the edges of the two graphene layers, which varies with the interlayer distance. This hybridization
affects the degree of localization of the electron orbitals near the edges, thus have a deciding role on the transport properties.

\subsection{\label{sec:31}Tunneling without overlapping between two graphene layers }

We first examine how the interlayer distance ($d_\perp$) and horizontal distance ($d_\parallel$) affect the tunneling properties when there is no overlap between the two sheets (see FIG. \ref{fig:dis}a). The spin-polarized calculations show that the edge of each graphene layer has a small amount of magnetization and the anti-ferromagnetic (AFM) bonding state between
the two edges is the ground state. The carbon atoms on equivalent positions from the two edges in each layer, e.g., C1 and C2, have the same magnitude of magnetization with opposite signs. Due to this spatially antisymmetric spin configuration, the transmission coefficient for the spin-down channel is identical to that for the spin-up channel \cite{spin2,spin1}.

With fixed  $d_\perp$, the calculated transmision as a function of  $d_\parallel$ shows an exponential decay, as plotted in FIG. \ref{fig:dis}b for $d_\perp=0$ and $d_\perp=2.1$ \AA .
Such a result is expected for simple tunneling through a vacuum barrier with a thickness equal to the separation $d_\parallel$.

The dependence on the interlayer distance ($d_\perp$) is more complex. We set $d_\parallel=0$ which yields the maximum transmission for non-overlapping sheets. In this case, the transmission coefficient first increases with the distance, reaching its maximum at $d_\perp=2.1$ \AA, followed by an exponential decrease for large distances (see FIG. \ref{fig:dis}c). The distance of $2.1$ \AA\ coincides with the minimum of the total absolute magnetization as shown in FIG. \ref{fig:dis}c, which also plots the total energy as a function of
$d_\perp$. The energy difference between ferromagnetic (FM) and AFM states is 5 $\,$meV when $d_\perp=1.5$ \AA . For other interlayer distances, we cannot find the FM states in the calculation. The equilibrium is reached when the interlayer distance is about $3.0$ \AA . The fact that the maximum transmission coincides with the minimum of absolute
magnetization suggests that there is a ``cancellation'' of the magnetic moments between the two opposing edges due to the majority spin electrons from each edge ``leaking'' into the
minority spin channel of the other side.

To understand the mechanism of this spin leakage, we examine the degree of the orbital hybridization between the edge atoms of the two graphene layers. In FIG. \ref{fig:pdos}, we plot the projected density of states (PDOS) of edge C1 and C2 atoms at several interlayer distances. When the interlayer distance is large (e.g. $d_\perp=4.0$ \AA , FIG. \ref{fig:pdos}a), C1 has a peak below the Fermi level for spin-up and a peak above the Fermi level for spin-down and C2 is the opposite, which gives them the same magnitude of magnetization but with opposite signs. When the distance becomes smaller (below around $3.0$ \AA ), we observe a small peak above the Fermi level in the spin-up PDOS of C1 (see FIG. \ref{fig:pdos}b), and correspondingly a small peak above the Fermi level in the spin-down PDOS of C2 (not shown here). This clearly indicates hybridization between the orbitals of C1 and C2. The degree of hybridization between the orbitals of the two graphene edges can be estimated from the size of the small peak in the spin-up PDOS of C1. The small peak reaches its maximum when the interlayer distance is $2.1$ \AA , indicating that the orbital hybridization is strongest at this distance, leading to the largest transmission coefficient. The size of the small peak is almost the same for $d_\perp=1.5$ \AA\ and $3.0$ \AA , consistent with Fig. \ref{fig:dis}c which shows that the transmission coefficient at these two interlayer distances is almost equal.

The degree of orbital hybridization as a function of $d_\perp$ can be visualized directly from the spin-up integrated local density of states (LDOS), as shown in FIG. \ref{fig:ldos}(a-d), which is calculated by integrating the spin-up DOS from $0.05$ eV below the incident energy to $0.05$ eV above the incident energy (in compliance with our smearing parameter). These plots show that the orbitals around incident energy that carry current are mainly the carbon $p_z$ orbitals. When $d_\perp$=4.0$\,\AA$ (FIG. \ref{fig:ldos}d), there is negligible LDOS on the top graphene layer (contains C1) indicating that there is almost no hybridization between the carbon orbitals from different graphene
sheets. This explains the exponentially decaying part of the transmission curve at large interlayer distance. With decreasing interlayer distance, one can observe some spin-up LDOS on the top graphene layer due to the hybridization between the spin-up orbitals of two carbon atoms at the edges of two sheets. Because of the hybridization, some electrons with the spin opposite to the magnetization
direction appear on the edge, as if they are ``leaked'' from the other edge whose magnetization is in the opposite direction, reducing the total absolute magnetization of the junction (see FIG. \ref{fig:dis}c).  The $p_z$ orbital in C1, which has the largest spin leakage compared to other carbon atoms on the top graphene layer, tilts with different angles at different interlayer distances, as shown in FIG. \ref{fig:ldos}e. At an angle when the orbital points directly at the C2 atom on the other edge, the hybridization, spin-leakage and degree of the LDOS delocalization in the top layer graphene are the largest. This happens at the interlayer distance of $2.1$ \AA\ (see FIG. \ref{fig:ldos}b). 

\subsection{\label{sec:34}Tunneling between two overlapping graphene layers}

In the case of two graphene layers overlapping each other, transmission depends sensitively on how the two layers are stacked together, and even whether there is a rotation angle between the two layers\cite{overlap}. There are two typical types of stacking. One, in which every atom on the second layer lies over an atom of the first, is called the AA
stacking. The other, in which half of the atoms in the second layer lie directly over the center of a hexagon in the lower sheet and the other half over an atom, is called the AB
stacking. We first examine the AA stacking pattern. We fix the interlayer distance of graphene layers to that of bilayer graphene, which is  about 3.4 $\AA$ \cite{interlayer}. By varying the overlapping area, expressed in the unit of graphene primitive unit cell area ($S_0$), we can calculate the transmission coefficient as a function of the overlapping area, which is plotted in FIG. \ref{fig:overlap}a for AA stacking. The transmission first increases superlinearly with the overlapping area, as it varies from two to eight graphene primitive cells. Then the increase slows for larger overlapping areas until the transmission coefficient appears to converge when the overlapping area exceeds ten graphene primitive cells. For large overlapping areas, electrode-limited conduction is reached, evident from  the transmission per conduction channel close to unity, a situation for which Eq. (\ref{T1_d}) is designed
to uncover. When the overlapping area is small, it presents a constriction to the current in the junction. The transmission per conduction channel of electrode is expected to be much smaller than unity in this case, and it is confirmed by our calculation. Small transmission due to the constriction at the junction will be referred to as  junction-limited transport. 

\begin{figure}[h]
{\includegraphics[width=0.5\textwidth]{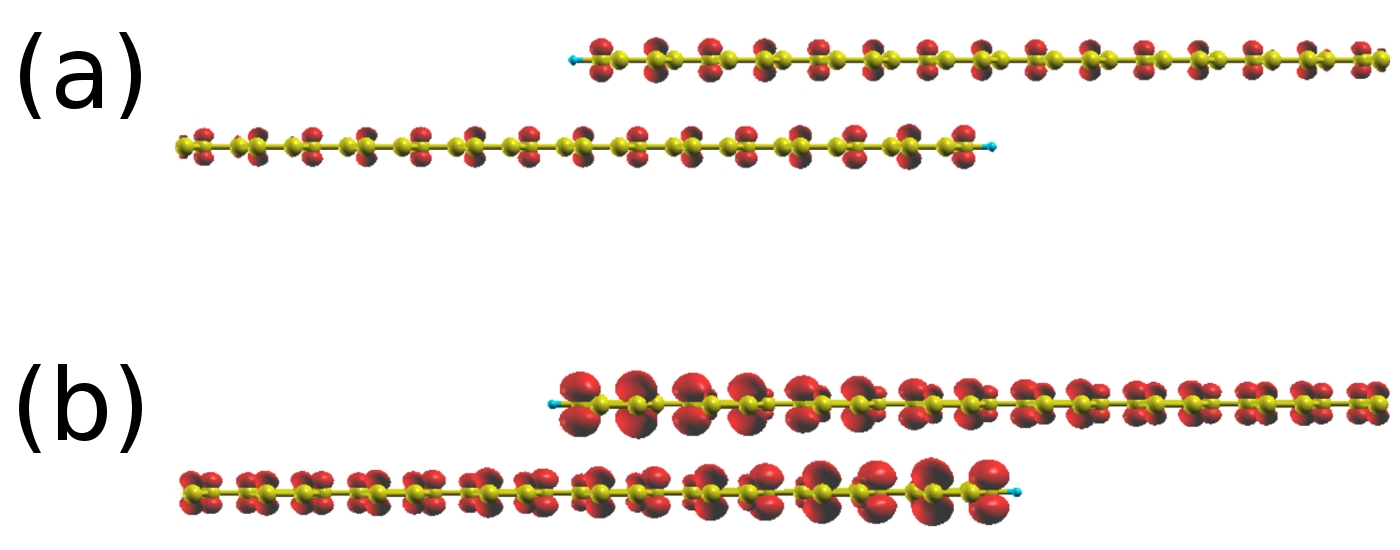}} \caption{\label{fig:ldosoverlap}  
(Color online) Isosurfaces (at 0.001 states/\AA$^3$, in red color) of integrated total local density of states (summed over both spins) from 0.05$\eV$ below the incident energy to 0.05$\eV$ above the incident energy for  \textbf{(a)} AB stacking with overlapping area $7S_0$, \textbf{(b)} AA stacking with overlapping area $8S_0$. $S_0$ is the area of graphene primitive unit cell.   Yellow balls are carbon atoms and blue balls are hydrogen atoms.}
\end{figure}

\begin{figure}[h]
{\includegraphics[width=0.5\textwidth]{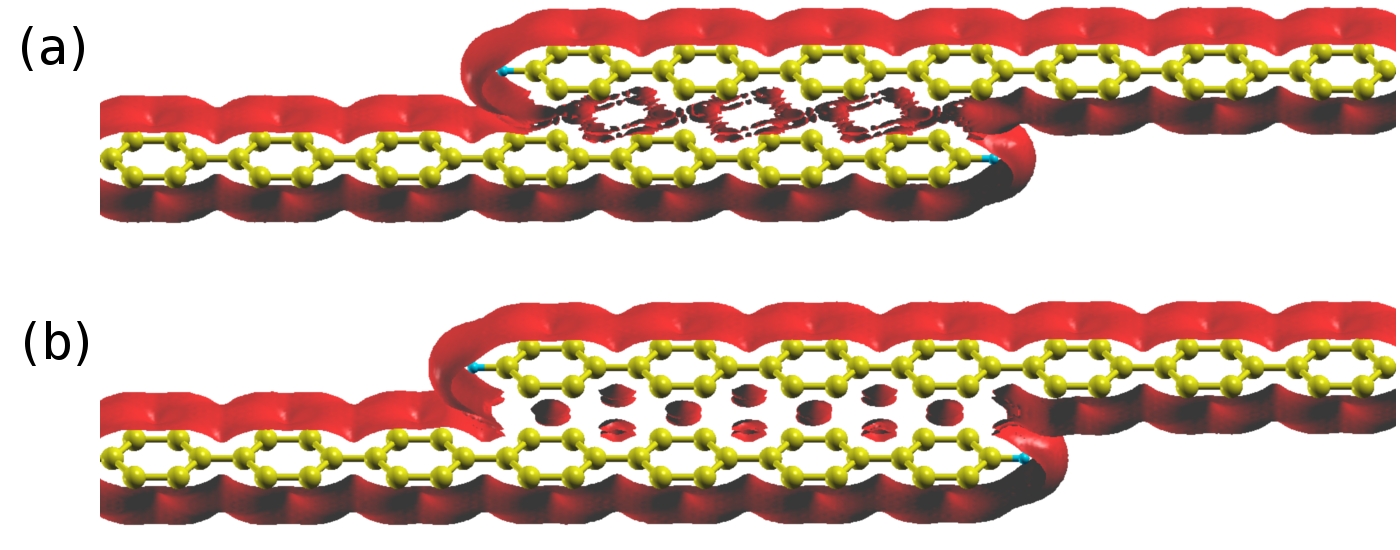}} \caption{\label{fig:pot}  
(Color online) Isosurface of the effective potential that electrons experience at the incident energy for  \textbf{(a)} AB stacking with overlapping area $7S_0$, \textbf{(b)} AA stacking with overlapping area $8S_0$. $S_0$ is the area of graphene primitive unit cell. The isosurface of the effective potential is in red. Yellow balls are carbon atoms and blue balls are hydrogen atoms.}
\end{figure}
\begin{figure}[h]
{\includegraphics[width=0.5\textwidth]{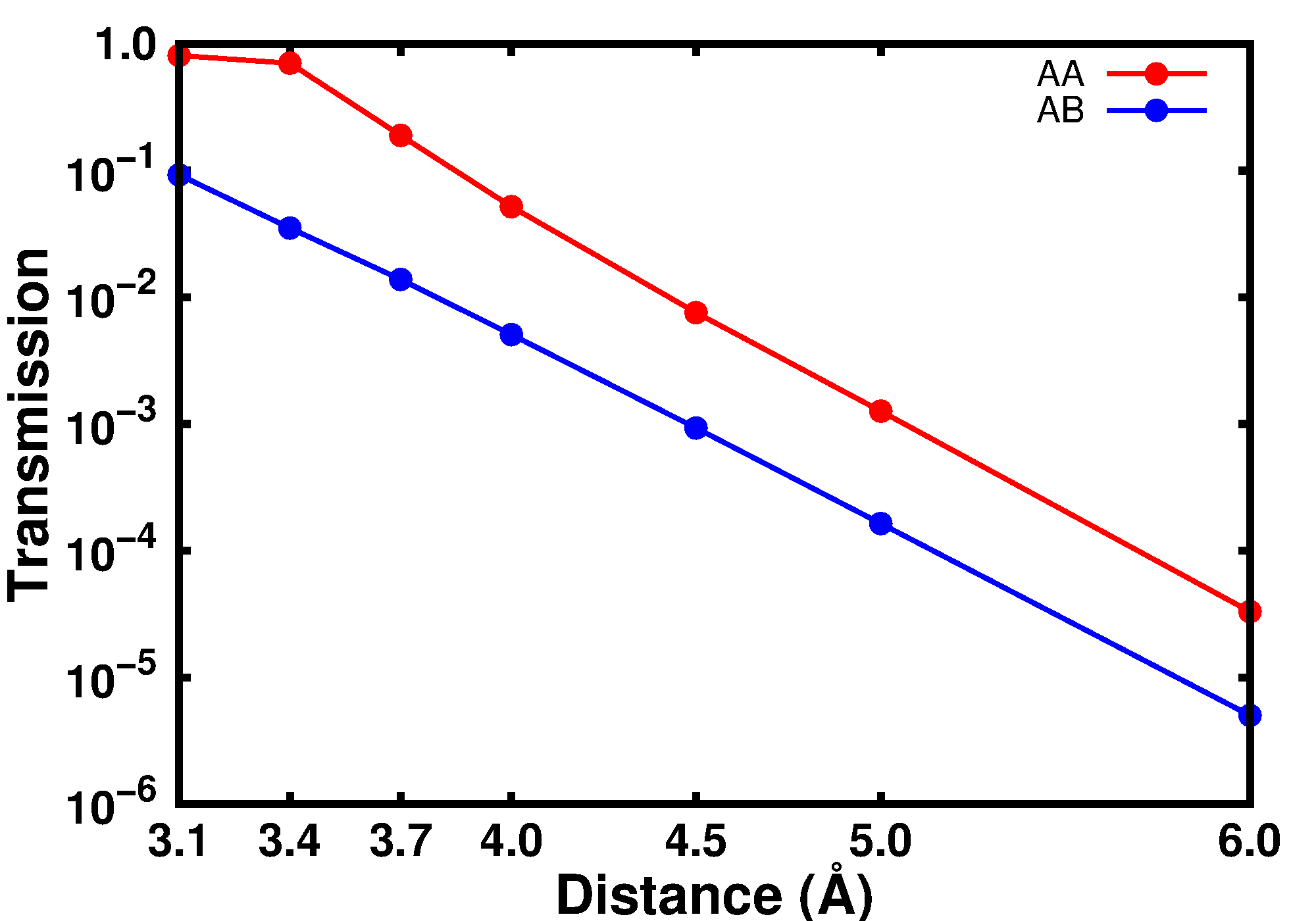}} \caption{\label{fig:T_d}  
(Color online) Transmission as a function of interlayer distance for both AA stacking with overlapping area $8S_0$ (red) and AB stacking with overlapping area $7S_0$ (blue). $S_0$ is the area of graphene primitive unit cell. $y$ axis is in logarithm scale.}
\end{figure}

For AB stacked junctions, the transmission as a function of overlapping area is shown in FIG.  \ref{fig:overlap}b. The transmission for AB stacking is about an order of magnitude smaller than that for AA stacking, as shown in FIG. \ref{fig:overlap}b. Consequently the transmission per conduction channel is much smaller than unity, placing AB stacked junctions within the junction-limited regime. The linear dependence of the transmission on the overlapping area depicted in FIG.  \ref{fig:overlap}b indicates a simple scaling of the tunneling
current with area. This may also be a consequence of weak coupling between the two layers, as evident from the much smaller
the total integrated LDOS of the carbon $p_z$ orbitals at the incident energy that carry the current for AB stacking than for  AA stacking, as shown in FIG. \ref{fig:ldosoverlap}.

To compare and relate the transport properties between AB and AA stacked graphene junctions, we examine the effective potential that electrons experience within the overlapping area between two graphene layers, plotted for the region outside the core radius, where the pseudopotential is equal to the all-electron potential. FIG. \ref{fig:pot} shows the isosurface of the effective potential at the incident energy (red color). The region of interest is between the two layers. The volume is divided by the red surface into regions with higher
potential than the incident energy, which are enclosed by the red surface and appear as solid red volumes, and regions with lower potential, which are the white colored regions between the two sheets (those outside both layers also have higher potentials).  In the case of AA stacking,   regions with higher potentials between two layers are contained within
isolated pockets, and regions with lower potentials form connected paths through the entire interlayer volume. For AB stacking, the white color low potential regions between the two layers are blocked off by the red color high potential regions. This difference in the topography of the potential can cause large difference in the transmission of the electron
wave function at the incident energy, leading to both differences in the conductance as well as wave function hybridization. 

We also plot in FIG. \ref{fig:T_d} the transmission coefficient as a function of the interlayer distance between the two layers for both AA and AB stacking. 
When the interlayer distance is larger than 4.5 $\,\AA$, both AA and AB stacking reach the vacuum tunneling regime giving us the same decaying rate. But the AB stacking reach the vacuum tunneling regime at a shorter interlayer distance than AA stacking, confirming that the interaction between the two layers is much smaller for AB stacking. 

\section{\label{sec:4}CONCLUSION}
By implementing the multiple scattering theory within the plane-wave basis, we have improved over previous plane-wave based
transport calculation in terms of both speed and parallel efficiency. We apply this method to study a graphene/graphene side contact junction system where the contact is formed by stacking two graphene layers through van der Waals interaction. The transmission through such a
junction is closely related to spin leakage between the two graphene edges, a consequence of orbital hybridization between the carbon
atoms across the layers which leads to the delocalization of the DOS. When the overlapping area is large, stacking pattern becomes an important factor in deciding transport properties across the layers. The transmission coefficients for AB stacking is one order of magnitude smaller than that for AA stacking, primarily due to the larger volume of the blocking potential within the overlapping region for AB stacking.

~

\centerline {\textbf{Acknowledgements}}

This work was supported by the US Department of Energy (DOE), Office of Basic Energy Sciences (BES), under Contract No.~DE-FG02-02ER45995. 
A portion of this research was conducted at the Center for Nanophase Materials Sciences, which is sponsored at Oak Ridge National Laboratory by the Division of Scientific User Facilities (X.-G. Z.). The computation was done using the utilities of the National Energy Research Scientific Computing Center (NERSC).
~
~
~
~

\onecolumngrid
{
\appendix

\section{Layer multiple scattering theory for nonlocal pseudopotentials}
The general solution of Eq. (\ref{kse0}) can be expressed as
\begin{eqnarray}
\psi({\bf r})=\sum_{n}a_{n}\psi_{n}({\bf r})+\sum_{\alpha lm}C_{\alpha lm}\psi_{\alpha lm}({\bf r}),\label{gsol}
\end{eqnarray}
where $\psi_{n}({\bf r})$ and $\psi_{\alpha lm}({\bf r})$ are the solutions of the following homogeneous and inhomogeneous equations, respectively,
\begin{eqnarray}
&&-\frac{\hbar^2}{2m}\nabla^2\psi_{n}({\bf r})+V_{loc}({\bf r})\psi_{n}({\bf r})=E\psi_{n}({\bf r}),\label{loceq}\\
&&-\frac{\hbar^2}{2m}\nabla^2\psi_{\alpha lm}({\bf r})+V_{loc}({\bf r})\psi_{\alpha lm}({\bf r})+\sum_{{\bf R}_{\bot}}e^{i{\bf k}_{\bot}\cdot{\bf R}_{\bot}}W_{lm}^{\alpha}({\bf r}-{\tau}^{\alpha}-{\bf R}_{\bot})=E\psi_{\alpha lm}({\bf r}).\label{nloceq}
\end{eqnarray}

We divide the system along the $z$-direction into $N$ slices. If the slices are sufficiently thin then within each slice $V_{loc}$ can be
approximated as independent of $z$. Applying Fourier transformation for the wavefunctions in the $p^{th}$ slice, we have,
\begin{eqnarray}
&& \psi^{p}_{n}({\bf r})=\phi^{p}_{n}({\bf r}_{\bot})e^{\pm ik^{p}_{n}(z-z_{p})}=\sum_{{\bf G}_{\bot}}\phi^{p}_{n}({\bf G}_{\bot})e^{i({\bf k}_{\bot}+{\bf G}_{\bot})\cdot{\bf r}_{\bot}}e^{\pm ik^{p}_{n}(z-z_{p})}\label{A6},\\
&& \psi^{p}_{\alpha lm}({\bf r})=\sum_{j}\phi^{p}_{j}({\bf r}_{\bot})f^{p}_{j,\alpha lm}(z)=\sum_{{\bf G}_{\bot}}\sum_{j}\phi^{p}_{j}({\bf G}_{\bot})e^{i({\bf k}_{\bot}+{\bf G}_{\bot})\cdot{\bf r}_{\bot}}f^{p}_{j,\alpha lm}(z)\label{A7},
\end{eqnarray}
where $\phi^{p}_{j}({\bf G}_{\bot})$ obeys the following eigenvalue equation,
\begin{eqnarray}
&& E^{p}_{j}\phi^{p}_{j}({\bf G}_{\bot})=\frac{\hbar^2}{2m}|{\bf k}_{\bot}+{\bf G}_{\bot}|^2\phi^{p}_{j}({\bf G}_{\bot})+\sum_{{\bf G}'_{\bot}}V_{loc}^{p}({\bf G}_{\bot}-{\bf G}'_{\bot})\phi^{p}_{j}({\bf G}'_{\bot})\label{A8},
\end{eqnarray}
and $f^{p}_{j,\alpha lm}(z)$ is given by,
\begin{eqnarray}
&& f^{p}_{j,\alpha lm}(z)=\sum_{{\bf G}_{\bot}}[\phi^{p}_{j}({\bf G}_{\bot})]^{*}e^{-i({\bf k}_{\bot}+{\bf G}_{\bot})\cdot\tau^{\alpha}_{\bot}}\int^{z_p}_{z_{p-1}}dz' {g^{p}_{j}(z-z')}W^{\alpha}_{lm}({\bf k}_{\bot}+{\bf G}_{\bot},z'-\tau^{\alpha}_{z})\label{A9},\\
&& W^{\alpha}_{lm}({\bf k}_{\bot}+{\bf G}_{\bot},z)=\frac{1}{\Omega_{2D}}\int d^2{\bf r}_{\bot}W^{\alpha}_{lm}({\bf r})e^{-i({\bf k}_{\bot}+{\bf G}_{\bot})\cdot{\bf r}_{\bot}}\label{A10}.
\end{eqnarray}
Here $\Omega_{2D}$ is the cross-sectional area of the two-dimensional supercell,  
$k^{p}_{j}=\sqrt{2m(E-E^{p}_{j})/\hbar}$ is the $z$-component of the wave vector in the $p^{th}$ slice, $V_{loc}^{p}({\bf G}_{\bot})$ is the Fourier transform of the local potential
in the $p^{th}$ slice, $g^{p}_{j}(z)=\exp(ik^{p}_{j}\cdot z)/2ik^{p}_{j}$ when $z>0$ and $g^{p}_{j}(z)=\exp(-ik^{p}_{j}\cdot z)/2ik^{p}_{j}$ when $z<0$. Calculations of $V_{p}({\bf G}_{\bot})$, $\phi^{p}_{j}({\bf G})$ and $f^{p}_{j,\alpha lm}(z)$ are similar to the Choi-Ihm method\cite{pwcond1}. 

The total wavefunction in the $p^{th}$ slice in Eq. (\ref{gsol_p0}) is obtained from,
\begin{eqnarray}
&& \psi^{p}_{\alpha lm}({\bf r})=\sum_{n}\phi^{p}_{n}({\bf r}_{\bot})f^{p}_{n,\alpha lm}(z),\\
&& C_{\alpha lm,k_i}=\sum_{uv}D_{lm,uv}\sum_{p=1}^{N}\int^{z_p}_{z_{p-1}}dz\int d^{2}{\bf r}_{\bot}[W^{\alpha}_{uv}({\bf r}-{\bf \tau}^{\alpha})]^{*}\psi^{p}_{k_i}({\bf r})\label{calm1}.
\end{eqnarray}

Matching the boundary conditions for the wavefunction and its derivative between any two adjacent slices, we obtain the recurrence relation for the expansion coefficients $A^{p}_{n,k_i}$ and $B^{p}_{n,k_i}$,
\begin{eqnarray}
A^{p}_{nk_i}&=&\frac{1}{2k_{n}^{p}}\left\{\sum_{j}A^{p+1}_{jk_i}\exp(-ik^{p+1}_{j}\Delta z)\left[(k^{p}_{n}+k^{p+1}_{j})\sum_{s}(M^{p+1}_{js})^{*}M^{p}_{ns}\right]
\right.\nonumber\\
&& \left.
+\sum_{j}B^{p+1}_{jk_i}\exp(ik^{p+1}_{j}\Delta z)\left[(k^{p}_{n}-k^{p+1}_{j})\sum_{s}(M^{p+1}_{js})^{*}M^{p}_{ns}\right ]\right\}+\sum_{\alpha lm}C_{\alpha lm,k_i}H^a_{n,\alpha lm}(p,p+1),\label{app}
\end{eqnarray}
\begin{eqnarray}
B^{p}_{nk_i}&=&\frac{1}{2k_{n}^p}\left\{\sum_{j}A^{p+1}_{jk_i}\exp(-ik^{p+1}_{j}\Delta z)\left[(k^{p}_{n}-k^{p+1}_{j})\sum_{s}(M^{p+1}_{js})^{*}M^{p}_{ns}\right ]
\right.\nonumber\\
&&  \left.
+\sum_{j}B^{p+1}_{jk_i}\exp(ik^{p+1}_{j}\Delta z)\left[(k^{p}_{n}+k^{p+1}_{j})\sum_{s}(M^{p+1}_{js})^{*}M^{p}_{ns}\right ]\right\}+\sum_{\alpha lm}C_{\alpha lm,k_i}H^b_{n,\alpha lm}(p,p+1),\label{bpp}
\end{eqnarray}
 where $H^a_{n,\alpha lm}(p,p+1)$, $H^b_{n,\alpha lm}(p,p+1)$ and $M^{p}_{ns}$ are, 
 \begin{eqnarray}
&& H^a_{n,\alpha lm}(p,p+1)\equiv\frac{1}{2k^p_{n}}\sum_{j}(k^p_{n}-k^{p+1}_{j})f^{p+1}_{j,\alpha lm}(z_p)\sum_{s}(M^{p+1}_{js})^*M^p_{ns}-f_{n,\alpha lm}^{p}(z_p),\label{Ha}\\
&&H^b_{n,\alpha lm}(p,p+1)\equiv\frac{1}{2k^p_{n}}\sum_{j}(k^p_{n}+k^{p+1}_{j})f^{p+1}_{j,\alpha lm}(z_p)\sum_{s}(M^{p+1}_{js})^{*}M^p_{ns},\label{Hb}\\
&& M^{p}_{ns}=\frac{1}{\Omega_{2D}}\int d{\bf r}_{\bot}
[\phi^p_{n}({\bf r}_{\bot})]^{*}\cdot\exp[i({\bf k}_{\bot}+{\bf G}_{\bot,s})\cdot{\bf r}_{\bot}],
\end{eqnarray}
and ${\bf G}_{\bot,s}$ denotes the $s^{th}$ reciprocal lattice vector in $\{{\bf G}_{\bot}\}$ and $\Delta z$ the thickness of a single slice along the $z$-direction.

The above recurrence relations can be rearranged to form an inhomogeneous equation,
\begin{eqnarray}
&&\left[ \begin{array}{c} A^{p} \\ B^{p} \end{array} \right] =\left[\begin{array}{c c} I_{11}(p,p+1)\ \ I_{12}(p,p+1)\\ I_{21}(p,p+1)\ \ I_{22}(p,p+1)\end{array}\right] \left[ \begin{array}{c} A^{p+1} \\ B^{p+1} \end{array} \right]+ \left[\begin{array}{c} H^{a}(p,p+1)C\\H^{b}(p,p+1)C\end{array} \right],\label{smm1}
\end{eqnarray}
where  $A^{p}$, $B^{p}$ and $C$ are matrices with elements $\{A^{p}_{nk_i}\}$, $\{B^{p}_{nk_i}\}$ and $\{C_{\alpha lm,k_i}\}$, respectively; $I(p,p+1)$ is the transfer matrix between adjacent slices $p$ and $p+1$ in the following form,
\begin{eqnarray}
&&I_{nj}(p,p+1) =\frac{1}{2k_{n}^{p}}\sum_{s}(M^{p+1}_{js})^{*}M^{p}_{ns}\left[\begin{array}{c c} \exp(-ik^{p+1}_{j}\Delta z)(k^{p}_{n}+k^{p+1}_{j}) \ \ \exp(ik^{p+1}_{j}\Delta z)(k^{p}_{n}-k^{p+1}_{j})\\ \exp(-ik^{p+1}_{j}\Delta z)(k^{p}_{n}-k^{p+1}_{j})\ \ \exp(ik^{p+1}_{j}\Delta z)(k^{p}_{n}+k^{p+1}_{j})\end{array}\right].\label{transfer}
\end{eqnarray}
The wave function is separated into forward and backward waves with the help of the scattering matrix, whose elements are calculated from the transfer matrix,
\begin{equation}
\begin{aligned}
S_{11}(p,p+1)&=[I_{11}(p,p+1)]^{-1},\\
S_{12}(p,p+1)&=-[I_{11}(p,p+1)]^{-1}I_{12}(p,p+1),\\
 S_{21}(p,p+1)&=I_{21}(p,p+1) S_{11}(p,p+1),\\
 S_{22}(p,p+1)&=I_{22}(p,p+1)+I_{21}(p,p+1) S_{12}(p,p+1),\\
 h^{a}(p,p+1)&=-[I_{11}(p,p+1)]^{-1} H^{a}(p,p+1),\\
 h^{b}(p,p+1)&=H^{b}(p,p+1)+I_{21}(p,p+1) h^{a}(p,p+1).\label{scatter}
\end{aligned}
\end{equation}
For $1\le m<n<k\le N$, we obtain the following recurrence relations,
\begin{equation}
\begin{aligned}
S_{11}(m,k)&=S_{11}(n,k)[I-S_{12}(m,n)S_{21}(n,k)]^{-1}S_{11}(m,n),\\
S_{12}(m,k)&=S_{12}(n,k)+S_{11}(n,k)[I-S_{12}(m,n)S_{21}(n,k)]^{-1}S_{12}(m,n)S_{22}(n,k),\\
S_{21}(m,k)&=S_{21}(m,n)+S_{22}(m,n)S_{21}(n,k)[I-S_{12}(m,n)S_{21}(n,k)]^{-1}S_{11}(m,n),\\
S_{22}(m,k)&=S_{22}(m,n)S_{22}(n,k)+S_{22}(m,n)S_{21}(n,k)[I-S_{12}(m,n)S_{21}(n,k)]^{-1}S_{12}(m,n)S_{22}(n,k),\\
h^{a}(m,k)&=h^{a}(n,k)+S_{11}(n,k)[I-S_{12}(m,n)S_{21}(n,k)]^{-1}[S_{12}(m,n)h^{b}(n,k)+h^{a}(m,n)],\\
h^{b}(m,k)&=h^{b}(m,n)+S_{22}(m,n)h^{b}(n,k)+S_{22}(m,n)S_{21}(n,k)[I-S_{12}(m,n)S_{21}(n,k)]^{-1}\times \\
&\quad [S_{12}(m,n)h^{b}(n,k)+h^{a}(m,n)].\label{recurrence}
\end{aligned}
\end{equation}
Each iteration of the recurrence relations results in the doubling of the number of slices represented by the the scattering matrix. This is called the
``doubling technique''.  At the end of the iteration, we obtain the scattering matrix that represent the entire region [see Eq. (\ref{smm3_0})]. To further
improve speed, we use a set of recurrence relations based on a mixture of transfer matrices $I_{11}$, $I_{12}$ and scattering matrices $S_{21}$, $S_{22}$ which is faster but numerically less stable,
\begin{equation}
\begin{aligned}
I_{11}(m,k)&=[I_{11}(m,n)+I_{12}(m,n)S_{21}(n,k)]I_{11}(n,k),\\
I_{12}(m,k)&=[I_{11}(m,n)+I_{12}(m,n)S_{21}(n,k)]I_{12}(n,k)+I_{12}(m,n)S_{22}(n,k),\\
S_{21}(m,k)&=S_{21}(m,n)+S_{22}(m,n)S_{21}(n,k)[I_{11}(m,n)+I_{12}(m,n)S_{21}(n,k)]^{-1},\\
S_{22}(m,k)&=S_{22}(m,n)S_{22}(n,k)-S_{22}(m,n)S_{21}(n,k)[I_{11}(m,n)+I_{12}(m,n)S_{21}(n,k)]^{-1}I_{12}(m,n)S_{22}(n,k),\\
H^{a}(m,k)&=[I_{11}(m,n)+I_{12}(m,n)S_{21}(n,k)]H^{a}(n,k)+I_{12}(m,n)h^{b}(n,k)+H^{a}(m,n),\\
h^{b}(m,k)&=h^{b}(m,n)+S_{22}(m,n)h^{b}(n,k)-S_{22}(m,n)S_{21}(n,k)[I_{11}(m,n)+I_{12}(m,n)S_{21}(n,k)]^{-1}\times \\
&\quad [I_{12}(m,n)h^{b}(n,k)+H^{a}(m,n)].\label{recurrence2}
\end{aligned}
\end{equation}
We iterate these equations until the condition number of $I_{11}$ exceeds a predetermined criterion. Then we switch to Eq. (\ref{recurrence}) after computing
$S_{11}$ and $S_{12}$ from $I_{11}$ and $I_{12}$ from Eq. (\ref{scatter}).

\section{Complex band structure calculation}
If the potential is periodic along the $z$-direction, the Bloch conditions can be applied as the boundary conditions,
\begin{eqnarray}
&& \psi_{k_i}({\bf r}_{\bot}+{\bf a}_{3\bot},z=d)=e^{i{\bf k}_{\bot}\cdot{\bf a}_{3\bot}+ikd}\psi_{k_i}({\bf r}_{\bot},z=0)\label{bc1},\\
&& \frac{\partial\psi_{k_i}({\bf r}_{\bot}+{\bf a}_{3\bot},z=d)}{\partial z}=e^{i{\bf k}_{\bot}\cdot{\bf a}_{3\bot}+ikd}\frac{\partial\psi_{k_i}({\bf r}_{\bot},z=0)}{\partial z}\label{bc2},
\end{eqnarray}
 where ${\bf a}_{3\bot}$ is the component of ${\bf a}_{3}$ in the $xy$ plane and $d$ is the $z$-component of ${\bf a}_{3}$. In a transport problem, the wave functions (of either electrode) do not extend to infinity in all directions -- they match to boundary conditions at the interfaces between the electrodes and the scattering region. Thus the requirement that the wave vector $k$ is real is
no longer necessary. Solutions with complex $k$ (evanescent waves) are now allowed, changing the above boundary conditions to the generalized Bloch conditions \cite{cmband}. States with real $k$'s are the propagating (Bloch) states and those with complex $k$'s are the evanescent states.
 
A complication of imposing the boundary conditions, Eqs. (\ref{bc1}) and (\ref{bc2}), on a nonlocal pseudopotential, is that one must account for the nonlocal
spheres that cross one boundary plane and are thus folded to the boundary plane on the other side of the supercell by the Bloch boundary condition. This requires  
Eq.(\ref{calm1}) to be rewritten in the following form when the nonlocal spheres (characterized by $W^{\alpha}_{lm}$) cross the left boundary of the unit cell at $z=0$, 
 \begin{eqnarray}
 C_{\alpha lm,k_i}=\sum_{uv}D_{lm,uv}\left[\int^{d}_{0}dz\int d^{2}{\bf r}_{\bot}[W^{\alpha}_{uv}({\bf r}-{\bf \tau}^{\alpha})]^{*}\psi_{k_i}({\bf r})+e^{-ikd}\int^{d}_{0}dz\int d^{2}{\bf r}_{\bot}[W^{\alpha}_{uv}({\bf r}-{\bf \tau}^{\alpha}-d\hat{z})]^{*}\psi_{k_i}({\bf r})\right]\label{calm2},
 \end{eqnarray}
  and  for those crossing the right boundary of the unit cell at $z=d$,
 \begin{eqnarray}
 C_{\alpha lm,k_i}=\sum_{uv}D_{lm,uv}\left[\int^{d}_{0}dz\int d^{2}{\bf r}_{\bot}[W^{\alpha}_{uv}({\bf r}-{\bf \tau}^{\alpha})]^{*}\psi_{k_i}({\bf r})+e^{ikd}\int^{d}_{0}dz\int d^{2}{\bf r}_{\bot}[W^{\alpha}_{uv}({\bf r}-{\bf \tau}^{\alpha}+d\hat{z})]^{*}\psi_{k_i}({\bf r})\right]\label{calm3}.
 \end{eqnarray}
 The total number of those spheres is $N_{crosl}+N_{crosr}$ with $N_{crosl} (N_{crosr})$ being the total number of nonlocal spheres crossing left (right) boundary of the unit cell. %

The unknowns in Eqs. (\ref{smm3_0}), (\ref{calm1}), (\ref{bc1}) and (\ref{bc2}) are $A^1$, $A^N$, $B^1$, $B^N$, $C_{(\alpha lm)^\prime}$ (nonlocal spheres completely fitting the electrode region) and $C_{(\alpha lm)}$ (nonlocal spheres crossing the boundaries of the electrode region). Some of these unknowns, $A^N$, $B^1$, and $C_{(\alpha lm)^\prime}$,  can be eliminated by expressing them in terms of $A^1$, $B^N$ and $C_{(\alpha lm)}$. The latter are collected as 
a single vector $X$,
\begin{eqnarray}
X =\left[\begin{array}{c } A_{nk_i}^1 \\\noalign{\smallskip} B_{nk_i}^N\\ C_{(\alpha lm), k_i}\end{array}\right].
\end{eqnarray}
Then the remaining equations from Eqs. (\ref{smm3_0}), (\ref{calm1}), (\ref{bc1}) and (\ref{bc2}) are combined into a generalized eigenvalue problem, Eq. (\ref{cmp}).
The dimension of $X$ is $(2N_{2D}+N_{crosl}+N_{crosr})\times$(number of incident waves).
$P$ and $Q$ in Eq. (\ref{cmp}) are $(2N_{2D}+N_{crosl}+N_{crosr})\times(2N_{2D}+N_{crosl}+N_{crosr})$ matrices.

Next, we describe how to fold the complex bands of each electrode from the bigger first Brillouin zone of the two-dimensional primitive cell, into the smaller Brillouin zone corresponding to the larger two-dimensional supercell of the junction system. Both the two-dimensional primitive cell and its corresponding supercell is in the transverse plane formed by the first two lattice vectors. The third lattice vector ${\bf a}_{3}$ is the same for both the primitive cell and the supercell. Let $\bf b_1$, $\bf b_2$ be the reciprocal lattice vectors of the electrode primitive cell in the plane perpendicular to ${\bf a}_3$ and $\bf B_1$, $\bf B_2$ the vectors for the supercell in the same plane. We know that the supercell's first two lattice vectors in real space can be expressed as a linear combination of the first two lattice vectors of the primitive cell with coefficients being integers, respectively. The inner product of the lattice vector in real space and its corresponding reciprocal lattice vector is a constant, $2\pi$. Thus we can write ${\bf b_1}=c_{11}{ \bf B_1}+c_{12}{ \bf B_2}$ and ${\bf b_2}=c_{21} {\bf B_1}+c_{22}{ \bf B_2}$, where $c_{11}$,  $c_{12}$, $c_{21}$ and $c_{22}$ are integers. We start from the electrode complex band structure described within the larger Brillouin zone of the primitive cell. For each $\bf k_{\bot}$ in the first brillouin zone, the Bloch condition along the third lattice vector ${\bf a}_{3}$ is,
\begin{eqnarray}
 \psi_{\bf k}({\bf r}+{\bf a}_{3})=e^{i{\bf k}\cdot{\bf a}_{3}}\psi_{\bf k}({\bf r})\label{bc5},
\end{eqnarray}
where ${\bf k}=({\bf k}_\bot,k_z)$.
The complex band structure calculation can give us a series of  eigenvalues $k_z$ and eigenstates $\psi_{k_z}$ [see Eq. (\ref{cmp})]. The Fourier transform of the electrode potential contains only vectors of the reciprocal lattice of the primitive cell, $m{\bf b_1}+n{\bf b_2}$, where m and n are integers. Thus each $\bf k_{\bot}$ in the first brillouin zone of the reciprocal space of the primitive cell couples only to ${\bf k}_{\bot}+m{\bf b_{1\bot}}+n{\bf b_{2\bot}}$. Then the eigenstate $\psi_{k_z}$ is superposition of plane waves containing only the wave vector $ {\bf k}_{\bot}$ and wave vectors differing from $ {\bf k}_{\bot}$ by $m{\bf b_{1\bot}}+n{\bf b_{2\bot}}$,
\begin{eqnarray}
\psi_{k_z}=\sum_m\sum_n a_{mnk}|{\bf k}_{\bot}+m{{\bf b}_{1\bot}}+n{{\bf b}_{2\bot}}\rangle,
\end{eqnarray} 
where $a_{mnk}$ are the expansion coefficients; $|{\bf k}_{\bot}+m{{\bf b}_{1\bot}}+n{{\bf b}_{2\bot}}\rangle$ is the plane wave basis.

 To fold ${\bf k}=({\bf k}_{\bot},k_z)$ of the primitive cell to ${\bf K}=({\bf K}_{\bot},K_z)$ of the supercell, we have the following relationship, 
\begin{eqnarray}
{\bf k }={\bf K}+m^k {\bf B_1} +n^k {\bf B_2},
\end{eqnarray}   
where $m^k$ and $n^k$ are integers. Then the Bloch condition along the third lattice vector ${\bf a}_{3}$ (supercell and the primitive cell have the same ${\bf a}_{3}$) for the supercell is,
\begin{eqnarray}
 \psi_{{\bf K}}({\bf r}+{\bf a}_{3})= \psi_{{\bf K}+m^k {\bf B_1} +n^k {\bf B_2}}({\bf r}+{\bf a}_{3})=\psi_{\bf k}({\bf r}+{\bf a}_{3})=e^{i{\bf k}\cdot{\bf a}_{3}}\psi_{\bf k}({\bf r})=e^{i({{\bf K}+m^k {\bf B_1} +n^k {\bf B_2}})\cdot{\bf a}_{3}}\psi_{{\bf K}+m^k {\bf B_1} +n^k {\bf B_2}}({\bf r})=e^{i{\bf K}\cdot{\bf a}_{3}}\psi_{\bf K}({\bf r})\label{bc5}.
\end{eqnarray}
Therefore, the complex band at $\bf k_\bot$ in the primitive cell will be folded to the complex band at $\bf K_{\bot}$ in the supercell. The eigenvalue $K_z=k_z-m^k {\bf B_{1z}} -n^k {\bf B_{2z}}$ and the corresponding eigenstate can be expanded in the new plane wave basis $|{\bf K}_{\bot}+M{\bf B_{1\bot}}+N{\bf B_{2\bot}}\rangle$, M and N are integers, as:
\begin{eqnarray}
\psi_{k_z}=\sum_m\sum_n a_{mnk}|{\bf k}_{\bot}+m{\bf b_{1\bot}}+n{\bf b_{2\bot}}\rangle=\sum_m\sum_n a_{mnk}|{\bf K}_{\bot}+(m^k+mc_{11}+nc_{21}){\bf B_{1\bot}}+(n^k+mc_{12}+nc_{22}){\bf B_{2\bot}}\rangle.
\end{eqnarray} 

\section{Reflection and transmission calculation}
 For transmission coefficient calculations, we need to match the boundary conditions at  $z_0=0$ and $z_{N}=d$ for the wavefunction and its derivative, which can be viewed as adding two more slices.  We denote the two added slices as the $0^{th}$ and the $(N+1)^{th}$ slices with wave coefficients $A^0$ and $B^0$, $A^{N+1}$ and $B^{N+1}$. Then we have
\begin{eqnarray}
A^{1}_{jk_i}&=&\frac{1}{2}\left\{\sum_{k_i^\prime \in R}A^{0}_{k_i^\prime k_i}\sum_{s}(M^{1}_{js})^{\dagger}\left[\psi_{k_i^\prime}({\bf G}_{\bot, s},z=0)-\frac{i}{k^1_j}\exp(ik_j^1\Delta z)\frac{\partial{\psi_{k_i^\prime}({\bf G}_{\bot, s},z=0)}}{\partial z}\right]
\right.\nonumber\\
&& \left.
+\sum_{k_i^\prime \in L}B^{0}_{k_i^\prime k_i}\sum_{s}(M^{1}_{js})^{\dagger}\left[\psi_{k_i^\prime}({\bf G}_{\bot, s},z=0)-\frac{i}{k^1_j}\exp(ik_j^1\Delta z)\frac{\partial{\psi_{k_i^\prime}({\bf G}_{\bot, s},z=0)}}{\partial z}\right]\right\}
\end{eqnarray}
\begin{eqnarray}
B^{1}_{jk_i}&=&\frac{1}{2}\left\{\sum_{k_i^\prime \in R}A^{0}_{k_i^\prime k_i}\sum_{s}(M^{1}_{js})^{\dagger}\left[\psi_{k_i^\prime}({\bf G}_{\bot, s},z=0)+\frac{i}{k^1_j}\exp(-ik_j^1\Delta z)\frac{\partial{\psi_{k_i^\prime}({\bf G}_{\bot, s},z=0)}}{\partial z}\right]
\right.\nonumber\\
&& \left.
+\sum_{k_i^\prime \in L}B^{0}_{k_i^\prime k_i}\sum_{s}(M^{1}_{js})^{\dagger}\left[\psi_{k_i^\prime}({\bf G}_{\bot, s},z=0)+\frac{i}{k^1_j}\exp(-ik_j^1\Delta z)\frac{\partial{\psi_{k_i^\prime}({\bf G}_{\bot, s},z=0)}}{\partial z}\right]\right\}\nonumber \\
&&+\sum_{\alpha lm}C_{\alpha lm,k_i}\left[-f^1_{j, \alpha lm}(0)\exp(-ik_j^1\Delta z)\right],
\end{eqnarray}
\begin{eqnarray}
A^{N}_{jk_i}&=&\frac{1}{2}\left\{\sum_{k_i^\prime \in R}A^{N+1}_{k_i^\prime k_i}\sum_{s}(M^{N}_{js})\left[\psi_{k_i^\prime}({\bf G}_{\bot, s},z=d)-\frac{i}{k^N_j}\frac{\partial{\psi_{k_i^\prime}({\bf G}_{\bot, s},z=d)}}{\partial z}\right]
\right.\nonumber\\
&& \left.
+\sum_{k_i^\prime \in L}B^{N+1}_{k_i^\prime k_i}\sum_{s}(M^{N}_{js})\left[\psi_{k_i^\prime}({\bf G}_{\bot, s},z=d)-\frac{i}{k^N_j}\frac{\partial{\psi_{k_i^\prime}({\bf G}_{\bot, s},z=d)}}{\partial z}\right]\right\}\nonumber \\
&&+\sum_{\alpha lm}C_{\alpha lm,k_i}\left[-f^N_{j, \alpha lm}(d)\right],
\end{eqnarray}
\begin{eqnarray}
B^{N}_{jk_i}&=&\frac{1}{2}\left\{\sum_{k_i^\prime \in R}A^{N+1}_{k_i^\prime k_i}\sum_{s}(M^{N}_{js})\left[\psi_{k_i^\prime}({\bf G}_{\bot, s},z=d)+\frac{i}{k^N_j}\frac{\partial{\psi_{k_i^\prime}({\bf G}_{\bot, s},z=d)}}{\partial z}\right]
\right.\nonumber\\
&& \left.
+\sum_{k_i^\prime \in L}B^{N+1}_{k_i^\prime k_i}\sum_{s}(M^{N}_{js})\left[\psi_{k_i^\prime}({\bf G}_{\bot, s},z=d)+\frac{i}{k^N_j}\frac{\partial{\psi_{k_i^\prime}({\bf G}_{\bot, s},z=d)}}{\partial z}\right]\right\},
\end{eqnarray}
where $L$ and $R$ represent the sets of forward and backward waves.
In the two new slices, we expand the wavefunction with the generalized Bloch basis (including both propagating waves and evanescent waves). Thus for each incident wave $k_i$, the dimensions of $A^0$, $B^0$ and $A^{N+1}$, $A^{N+1}$ are $N_{2D}+N_{crosl}$ and $N_{2D}+N_{crosr}$, respectively. Recall that the dimensions of $A^1$, $B^1$ and $A^{N}$, $A^{N}$ are all $N_{2D}$. Thus we rearrange the above equations into the following form,
\begin{eqnarray}
\left[ \begin{array}{c} A^{1} \\ B^{1} \end{array} \right] &=&\left[\begin{array}{c c} I_{11}(1,0)\ \ I_{12}(1,0)\\ I_{21}(1,0)\ \ I_{22}(1,0)\end{array}\right] \left[ \begin{array}{c} A^{0} \\ B^{0} \end{array} \right]+ \left[\begin{array}{c} H^{a}(1,0)C\\H^{b}(1,0)C\end{array} \right],\label{left}
\end{eqnarray}
\begin{eqnarray}
\left[ \begin{array}{c} A^{N} \\ B^{N} \end{array} \right] &=&\left[\begin{array}{c c} I_{11}(N,N+1)\ \ I_{12}(N,N+1)\\ I_{21}(N,N+1)\ \ I_{22}(N,N+1)\end{array}\right] \left[ \begin{array}{c} A^{N+1} \\ B^{N+1} \end{array} \right]+ \left[\begin{array}{c} H^{a}(N,N+1)C\\H^{b}(N,N+1)C\end{array} \right],\label{right}
\end{eqnarray}
where
 \begin{eqnarray}
&& H^a_{j,\alpha lm}(1,0)=0,\\
&& H^b_{j,\alpha lm}(1,0)=-f^1_{j, \alpha lm}(0)\exp(-ik_j^1\Delta z),\\
&& H^a_{j,\alpha lm}(N,N+1)=-f^N_{j, \alpha lm}(d),\\
&& H^b_{j,\alpha lm}(N,N+1)=0,
\end{eqnarray}
\begin{eqnarray}
I_{jk_i^\prime}(1,0)&=&\frac{1}{2}\sum_{s}(M^{1}_{js})^{\dagger}\left[\begin{array}{c}\psi_{k_i^\prime}({\bf G}_{\bot, s},z=0)-\frac{i}{k^1_j}\exp(ik_j^1\Delta z)\frac{\partial{\psi_{k_i^\prime}({\bf G}_{\bot, s},z=0)}}{\partial z} \\ \psi_{k_i^\prime}({\bf G}_{\bot, s},z=0)+\frac{i}{k^1_j}exp(-ik_j^1\Delta z)\frac{\partial{\psi_{k_i^\prime}({\bf G}_{\bot, s},z=0)}}{\partial z}\end{array}\right],
\end{eqnarray}
\begin{eqnarray}
I_{jk_i^\prime}(N,N+1)&=&\frac{1}{2}\sum_{s}(M^{N}_{js})\left[\begin{array}{c}\psi_{k_i^\prime}({\bf G}_{\bot, s},z=d)-\frac{i}{k^N_j}\frac{\partial{\psi_{k_i^\prime}({\bf G}_{\bot, s},z=d)}}{\partial z} \\ \psi_{k_i^\prime}({\bf G}_{\bot, s},z=d)+\frac{i}{k^N_j}\frac{\partial{\psi_{k_i^\prime}({\bf G}_{\bot, s},z=d)}}{\partial z}\end{array}\right].
\end{eqnarray}
In addition, Eq. (\ref{smm3_0}) can be rewritten as 
\begin{eqnarray}
&&\left[\begin{array}{cc} I \ \ -S_{12}(1,N)\\ 0 \ \ -S_{22}(1,N)\end{array}\right]\left[ \begin{array}{c} A^{N} \\ B^{N} \end{array} \right] =\left[\begin{array}{c c} S_{11}(1,N) \ \ 0 \\ S_{21}(1,N) \ \ -I \end{array}\right] \left[ \begin{array}{c} A^{1} \\ B^{1} \end{array} \right]+ \left[\begin{array}{c} h^{a}(1,N)C\\h^{b}(1,N)C\end{array} \right].\label{new} 
\end{eqnarray}
From Eqns. (\ref{left}), (\ref{right}) and (\ref{new}), we can construct the relationship between $A^0$, $B^0$ and $A^{N+1}$, $A^{N+1}$,
\begin{eqnarray}
\left[\begin{array}{cc} I \ \ -S_{12}(1,N)\\ 0 \ \ -S_{22}(1,N)\end{array}\right]\left[\begin{array}{c c} I_{11}(N,N+1)\ \ I_{12}(N,N+1)\\ I_{21}(N,N+1)\ \ I_{22}(N,N+1)\end{array}\right] \left[ \begin{array}{c} A^{N+1} \\ B^{N+1} \end{array} \right]=
\left[\begin{array}{c c} S_{11}(1,N) \ \ 0 \\ S_{21}(1,N) \ \ -I \end{array}\right] \times &&\nonumber \\  
\left[\begin{array}{c c} I_{11}(1,0)\ \ I_{12}(1,0)\\ I_{21}(1,0)\ \ I_{22}(1,0)\end{array}\right] \left[ \begin{array}{c} A^{0} \\ B^{0} \end{array} \right]+ \left[\begin{array}{c} H^{a}(1,0)+ h^{a}(1,N)-H^{a}(N,N+1)\\H^{b}(1,0)+ h^{b}(1,N)-H^{b}(N,N+1)\end{array} \right]C &&.\label{bound}
\end{eqnarray}
Because the left electrode and the scattering region share those atoms whose nonlocal spheres lie across the $z=0$ plane, we have $N_{crosl}$ additional equations in the form,
\begin{eqnarray}
C_{(\alpha lm),k_i}=\sum_{k^{\prime}_i\in R}C_{(\alpha lm),k^{\prime}_i}^{LE}A^{0}_{k^{\prime}_i,k_i}+\sum_{k^{\prime}_i\in L}C_{(\alpha lm),k^{\prime}_i}^{LE}B^{0}_{k^{\prime}_i,k_i}.\label{C2}
\end{eqnarray}
Similarly, for those spheres lying across the $z=d$ plane, we have $N_{crosr}$ additional equations
\begin{eqnarray}
C_{(\alpha lm),k_i}=\sum_{k^{\prime}_i\in R}C_{(\alpha lm),k^{\prime}_i}^{RE}A^{N+1}_{k^{\prime}_i,k_i}+\sum_{k^{\prime}_i\in L}C_{(\alpha lm),k^{\prime}_i}^{RE}B^{N+1}_{k^{\prime}_i,k_i}.\label{C3}
\end{eqnarray}
Here $LE$ ($RE$) means left (right) electrode. 

Therefore for each boundary conditon,  altogether we have $2N_{2D}+N_{orb}+N_{crosl}+N_{crosr}$ equations from Eqns. (\ref{bound}) ($2N_{2D}$), (\ref{C2}) ($N_{crosl}$), (\ref{C3}) ($N_{crosr}$), (\ref{calm1}) ($N_{orb}$). The number of unknowns for $\{A^{N+1}\}$, $\{B^{0}\}$ and $\{C\}$ are $\{N_{2D}+N_{crosl}\}+\{N_{2D}+N_{crosr}\}+\{N_{orb}\}$. Note that $\{A^{0}\}$ and $\{B^{N+1}\}$ provide the boundary conditions that specify the incident wave, e.g. $A^{0}={ I}$, $B^{N+1}={ 0}$ for waves incident from the left electrode for which the corresponding transmission and reflection matrices are $\{A^{N+1}\}$ and $\{B^{0}\}$, respectively. $A^{0}={ 0}$, $B^{N+1}={ I}$ represents waves incident from the right electrode for which  the corresponding transmission and reflection matrices are $\{B^{0}\}$ and $\{A^{N+1}\}$, respectively.
As explained in the main text, rearrange Eqs.(\ref{bound}), (\ref{C2}), (\ref{C3}) and (\ref{calm1}),  we can obtain a set of linear equations listed in Eq. (\ref{trans}), where $M$ is a $(2N_{2D}+N_{crosl}+N_{crosr}+N_{orb})$$\times$$(2N_{2D}+N_{crosl}+N_{crosr}+N_{orb})$ matrix, $D$ is $(2N_{2D}+N_{crosl}+N_{crosr}+N_{orb})$$\times$(number of incident waves), $X$ is also $(2N_{2D}+N_{crosl}+N_{crosr}+N_{orb})$$\times$(number of incident waves) and has the following structure,
\begin{eqnarray}
X =\left[\begin{array}{c } A^{N+1} \\ B^0\\ C\end{array}\right].
\end{eqnarray}

\twocolumngrid

\bibliography{bibtex}

\begin{thebibliography}{51}%
\makeatletter
\providecommand \@ifxundefined [1]{%
 \@ifx{#1\undefined}
}%
\providecommand \@ifnum [1]{%
 \ifnum #1\expandafter \@firstoftwo
 \else \expandafter \@secondoftwo
 \fi
}%
\providecommand \@ifx [1]{%
 \ifx #1\expandafter \@firstoftwo
 \else \expandafter \@secondoftwo
 \fi
}%
\providecommand \natexlab [1]{#1}%
\providecommand \enquote  [1]{``#1''}%
\providecommand \bibnamefont  [1]{#1}%
\providecommand \bibfnamefont [1]{#1}%
\providecommand \citenamefont [1]{#1}%
\providecommand \href@noop [0]{\@secondoftwo}%
\providecommand \href [0]{\begingroup \@sanitize@url \@href}%
\providecommand \@href[1]{\@@startlink{#1}\@@href}%
\providecommand \@@href[1]{\endgroup#1\@@endlink}%
\providecommand \@sanitize@url [0]{\catcode `\\12\catcode `\$12\catcode
  `\&12\catcode `\#12\catcode `\^12\catcode `\_12\catcode `\%12\relax}%
\providecommand \@@startlink[1]{}%
\providecommand \@@endlink[0]{}%
\providecommand \url  [0]{\begingroup\@sanitize@url \@url }%
\providecommand \@url [1]{\endgroup\@href {#1}{\urlprefix }}%
\providecommand \urlprefix  [0]{URL }%
\providecommand \Eprint [0]{\href }%
\providecommand \doibase [0]{http://dx.doi.org/}%
\providecommand \selectlanguage [0]{\@gobble}%
\providecommand \bibinfo  [0]{\@secondoftwo}%
\providecommand \bibfield  [0]{\@secondoftwo}%
\providecommand \translation [1]{[#1]}%
\providecommand \BibitemOpen [0]{}%
\providecommand \bibitemStop [0]{}%
\providecommand \bibitemNoStop [0]{.\EOS\space}%
\providecommand \EOS [0]{\spacefactor3000\relax}%
\providecommand \BibitemShut  [1]{\csname bibitem#1\endcsname}%
\let\auto@bib@innerbib\@empty
\bibitem [{\citenamefont {Geim}\ and\ \citenamefont {Novoselov}(2007)}]{gra1}%
  \BibitemOpen
  \bibfield  {author} {\bibinfo {author} {\bibfnamefont {A.~K.}\ \bibnamefont
  {Geim}}\ and\ \bibinfo {author} {\bibfnamefont {K.~S.}\ \bibnamefont
  {Novoselov}},\ }\href {\doibase 10.1038/nmat1849} {\bibfield  {journal}
  {\bibinfo  {journal} {Nat Mater.}\ }\textbf {\bibinfo {volume} {6}},\
  \bibinfo {pages} {183} (\bibinfo {year} {2007})}\BibitemShut {NoStop}%
\bibitem [{\citenamefont {Geim}(2009)}]{gra2}%
  \BibitemOpen
  \bibfield  {author} {\bibinfo {author} {\bibfnamefont {A.~K.}\ \bibnamefont
  {Geim}},\ }\href {\doibase 10.1126/science.1158877} {\bibfield  {journal}
  {\bibinfo  {journal} {Science}\ }\textbf {\bibinfo {volume} {324}},\ \bibinfo
  {pages} {1530} (\bibinfo {year} {2009})}\BibitemShut {NoStop}%
\bibitem [{\citenamefont {Huang}\ \emph {et~al.}(2011)\citenamefont {Huang},
  \citenamefont {Ruiz-Vargas}, \citenamefont {van~der Zande}, \citenamefont
  {Whitney}, \citenamefont {Levendorf}, \citenamefont {Kevek}, \citenamefont
  {Garg}, \citenamefont {Alden}, \citenamefont {Hustedt}, \citenamefont {Zhu},
  \citenamefont {Park}, \citenamefont {McEuen},\ and\ \citenamefont
  {Muller}}]{grain1}%
  \BibitemOpen
  \bibfield  {author} {\bibinfo {author} {\bibfnamefont {P.~Y.}\ \bibnamefont
  {Huang}}, \bibinfo {author} {\bibfnamefont {C.~S.}\ \bibnamefont
  {Ruiz-Vargas}}, \bibinfo {author} {\bibfnamefont {A.~M.}\ \bibnamefont
  {van~der Zande}}, \bibinfo {author} {\bibfnamefont {W.~S.}\ \bibnamefont
  {Whitney}}, \bibinfo {author} {\bibfnamefont {M.~P.}\ \bibnamefont
  {Levendorf}}, \bibinfo {author} {\bibfnamefont {J.~W.}\ \bibnamefont
  {Kevek}}, \bibinfo {author} {\bibfnamefont {S.}~\bibnamefont {Garg}},
  \bibinfo {author} {\bibfnamefont {J.~S.}\ \bibnamefont {Alden}}, \bibinfo
  {author} {\bibfnamefont {C.~J.}\ \bibnamefont {Hustedt}}, \bibinfo {author}
  {\bibfnamefont {Y.}~\bibnamefont {Zhu}}, \bibinfo {author} {\bibfnamefont
  {J.}~\bibnamefont {Park}}, \bibinfo {author} {\bibfnamefont {P.~L.}\
  \bibnamefont {McEuen}}, \ and\ \bibinfo {author} {\bibfnamefont {D.~A.}\
  \bibnamefont {Muller}},\ }\href {\doibase 10.1038/nature09718} {\bibfield
  {journal} {\bibinfo  {journal} {Nature.}\ }\textbf {\bibinfo {volume}
  {469}},\ \bibinfo {pages} {389} (\bibinfo {year} {2011})}\BibitemShut
  {NoStop}%
\bibitem [{\citenamefont {Kim}\ \emph {et~al.}(2011)\citenamefont {Kim},
  \citenamefont {Lee}, \citenamefont {Regan}, \citenamefont {Kisielowski},
  \citenamefont {Crommie},\ and\ \citenamefont {Zettl}}]{grain2}%
  \BibitemOpen
  \bibfield  {author} {\bibinfo {author} {\bibfnamefont {K.}~\bibnamefont
  {Kim}}, \bibinfo {author} {\bibfnamefont {Z.}~\bibnamefont {Lee}}, \bibinfo
  {author} {\bibfnamefont {W.}~\bibnamefont {Regan}}, \bibinfo {author}
  {\bibfnamefont {C.}~\bibnamefont {Kisielowski}}, \bibinfo {author}
  {\bibfnamefont {M.~F.}\ \bibnamefont {Crommie}}, \ and\ \bibinfo {author}
  {\bibfnamefont {A.}~\bibnamefont {Zettl}},\ }\href {\doibase
  10.1021/nn1033423} {\bibfield  {journal} {\bibinfo  {journal} {ACS Nano}\
  }\textbf {\bibinfo {volume} {5}},\ \bibinfo {pages} {2142} (\bibinfo {year}
  {2011})}\BibitemShut {NoStop}%
\bibitem [{\citenamefont {Li}\ \emph {et~al.}(2010)\citenamefont {Li},
  \citenamefont {Magnuson}, \citenamefont {Venugopal}, \citenamefont {An},
  \citenamefont {Suk}, \citenamefont {Han}, \citenamefont {Borysiak},
  \citenamefont {Cai}, \citenamefont {Velamakanni}, \citenamefont {Zhu},
  \citenamefont {Fu}, \citenamefont {Vogel}, \citenamefont {Voelkl},
  \citenamefont {Colombo},\ and\ \citenamefont {Ruoff}}]{grain3}%
  \BibitemOpen
  \bibfield  {author} {\bibinfo {author} {\bibfnamefont {X.}~\bibnamefont
  {Li}}, \bibinfo {author} {\bibfnamefont {C.~W.}\ \bibnamefont {Magnuson}},
  \bibinfo {author} {\bibfnamefont {A.}~\bibnamefont {Venugopal}}, \bibinfo
  {author} {\bibfnamefont {J.}~\bibnamefont {An}}, \bibinfo {author}
  {\bibfnamefont {J.~W.}\ \bibnamefont {Suk}}, \bibinfo {author} {\bibfnamefont
  {B.}~\bibnamefont {Han}}, \bibinfo {author} {\bibfnamefont {M.}~\bibnamefont
  {Borysiak}}, \bibinfo {author} {\bibfnamefont {W.}~\bibnamefont {Cai}},
  \bibinfo {author} {\bibfnamefont {A.}~\bibnamefont {Velamakanni}}, \bibinfo
  {author} {\bibfnamefont {Y.}~\bibnamefont {Zhu}}, \bibinfo {author}
  {\bibfnamefont {L.}~\bibnamefont {Fu}}, \bibinfo {author} {\bibfnamefont
  {E.~M.}\ \bibnamefont {Vogel}}, \bibinfo {author} {\bibfnamefont
  {E.}~\bibnamefont {Voelkl}}, \bibinfo {author} {\bibfnamefont
  {L.}~\bibnamefont {Colombo}}, \ and\ \bibinfo {author} {\bibfnamefont
  {R.~S.}\ \bibnamefont {Ruoff}},\ }\href {\doibase 10.1021/nl101629g}
  {\bibfield  {journal} {\bibinfo  {journal} {Nano Letters}\ }\textbf {\bibinfo
  {volume} {10}},\ \bibinfo {pages} {4328} (\bibinfo {year}
  {2010})}\BibitemShut {NoStop}%
\bibitem [{\citenamefont {Liu}\ and\ \citenamefont {Yakobson}(2010)}]{grain4}%
  \BibitemOpen
  \bibfield  {author} {\bibinfo {author} {\bibfnamefont {Y.}~\bibnamefont
  {Liu}}\ and\ \bibinfo {author} {\bibfnamefont {B.~I.}\ \bibnamefont
  {Yakobson}},\ }\href {\doibase 10.1021/nl100988r} {\bibfield  {journal}
  {\bibinfo  {journal} {Nano Letters}\ }\textbf {\bibinfo {volume} {10}},\
  \bibinfo {pages} {2178} (\bibinfo {year} {2010})}\BibitemShut {NoStop}%
\bibitem [{\citenamefont {Yazyev}\ and\ \citenamefont
  {Louie}(2010{\natexlab{a}})}]{study1}%
  \BibitemOpen
  \bibfield  {author} {\bibinfo {author} {\bibfnamefont {O.~V.}\ \bibnamefont
  {Yazyev}}\ and\ \bibinfo {author} {\bibfnamefont {S.~G.}\ \bibnamefont
  {Louie}},\ }\href {\doibase 10.1038/nmat2830} {\bibfield  {journal} {\bibinfo
   {journal} {Nat Mater.}\ }\textbf {\bibinfo {volume} {9}},\ \bibinfo {pages}
  {806} (\bibinfo {year} {2010}{\natexlab{a}})}\BibitemShut {NoStop}%
\bibitem [{\citenamefont {Bagri}\ \emph {et~al.}(2011)\citenamefont {Bagri},
  \citenamefont {Kim}, \citenamefont {Ruoff},\ and\ \citenamefont
  {Shenoy}}]{study2}%
  \BibitemOpen
  \bibfield  {author} {\bibinfo {author} {\bibfnamefont {A.}~\bibnamefont
  {Bagri}}, \bibinfo {author} {\bibfnamefont {S.-P.}\ \bibnamefont {Kim}},
  \bibinfo {author} {\bibfnamefont {R.~S.}\ \bibnamefont {Ruoff}}, \ and\
  \bibinfo {author} {\bibfnamefont {V.~B.}\ \bibnamefont {Shenoy}},\ }\href
  {\doibase 10.1021/nl202118d} {\bibfield  {journal} {\bibinfo  {journal} {Nano
  Letters}\ }\textbf {\bibinfo {volume} {11}},\ \bibinfo {pages} {3917}
  (\bibinfo {year} {2011})}\BibitemShut {NoStop}%
\bibitem [{\citenamefont {Jauregui}\ \emph {et~al.}(2011)\citenamefont
  {Jauregui}, \citenamefont {Cao}, \citenamefont {Wu}, \citenamefont {Yu},\
  and\ \citenamefont {Chen}}]{study3}%
  \BibitemOpen
  \bibfield  {author} {\bibinfo {author} {\bibfnamefont {L.~A.}\ \bibnamefont
  {Jauregui}}, \bibinfo {author} {\bibfnamefont {H.}~\bibnamefont {Cao}},
  \bibinfo {author} {\bibfnamefont {W.}~\bibnamefont {Wu}}, \bibinfo {author}
  {\bibfnamefont {Q.}~\bibnamefont {Yu}}, \ and\ \bibinfo {author}
  {\bibfnamefont {Y.~P.}\ \bibnamefont {Chen}},\ }\href {\doibase
  http://dx.doi.org/10.1016/j.ssc.2011.05.023} {\bibfield  {journal} {\bibinfo
  {journal} {Solid State Communications}\ }\textbf {\bibinfo {volume} {151}},\
  \bibinfo {pages} {1100 } (\bibinfo {year} {2011})}\BibitemShut {NoStop}%
\bibitem [{\citenamefont {Tsen}\ \emph {et~al.}(2012)\citenamefont {Tsen},
  \citenamefont {Brown}, \citenamefont {Levendorf}, \citenamefont {Ghahari},
  \citenamefont {Huang}, \citenamefont {Havener}, \citenamefont {Ruiz-Vargas},
  \citenamefont {Muller}, \citenamefont {Kim},\ and\ \citenamefont
  {Park}}]{study4}%
  \BibitemOpen
  \bibfield  {author} {\bibinfo {author} {\bibfnamefont {A.~W.}\ \bibnamefont
  {Tsen}}, \bibinfo {author} {\bibfnamefont {L.}~\bibnamefont {Brown}},
  \bibinfo {author} {\bibfnamefont {M.~P.}\ \bibnamefont {Levendorf}}, \bibinfo
  {author} {\bibfnamefont {F.}~\bibnamefont {Ghahari}}, \bibinfo {author}
  {\bibfnamefont {P.~Y.}\ \bibnamefont {Huang}}, \bibinfo {author}
  {\bibfnamefont {R.~W.}\ \bibnamefont {Havener}}, \bibinfo {author}
  {\bibfnamefont {C.~S.}\ \bibnamefont {Ruiz-Vargas}}, \bibinfo {author}
  {\bibfnamefont {D.~A.}\ \bibnamefont {Muller}}, \bibinfo {author}
  {\bibfnamefont {P.}~\bibnamefont {Kim}}, \ and\ \bibinfo {author}
  {\bibfnamefont {J.}~\bibnamefont {Park}},\ }\href {\doibase
  10.1126/science.1218948} {\bibfield  {journal} {\bibinfo  {journal}
  {Science}\ }\textbf {\bibinfo {volume} {336}},\ \bibinfo {pages} {1143}
  (\bibinfo {year} {2012})}\BibitemShut {NoStop}%
\bibitem [{\citenamefont {Yazyev}\ and\ \citenamefont
  {Louie}(2010{\natexlab{b}})}]{study5}%
  \BibitemOpen
  \bibfield  {author} {\bibinfo {author} {\bibfnamefont {O.~V.}\ \bibnamefont
  {Yazyev}}\ and\ \bibinfo {author} {\bibfnamefont {S.~G.}\ \bibnamefont
  {Louie}},\ }\href {\doibase 10.1103/PhysRevB.81.195420} {\bibfield  {journal}
  {\bibinfo  {journal} {Phys. Rev. B}\ }\textbf {\bibinfo {volume} {81}},\
  \bibinfo {pages} {195420} (\bibinfo {year} {2010}{\natexlab{b}})}\BibitemShut
  {NoStop}%
\bibitem [{\citenamefont {Robertson}\ \emph {et~al.}(2011)\citenamefont
  {Robertson}, \citenamefont {Bachmatiuk}, \citenamefont {Wu}, \citenamefont
  {Schäffel}, \citenamefont {Rellinghaus}, \citenamefont {Büchner},
  \citenamefont {Rümmeli},\ and\ \citenamefont {Warner}}]{overlap}%
  \BibitemOpen
  \bibfield  {author} {\bibinfo {author} {\bibfnamefont {A.~W.}\ \bibnamefont
  {Robertson}}, \bibinfo {author} {\bibfnamefont {A.}~\bibnamefont
  {Bachmatiuk}}, \bibinfo {author} {\bibfnamefont {Y.~A.}\ \bibnamefont {Wu}},
  \bibinfo {author} {\bibfnamefont {F.}~\bibnamefont {Schäffel}}, \bibinfo
  {author} {\bibfnamefont {B.}~\bibnamefont {Rellinghaus}}, \bibinfo {author}
  {\bibfnamefont {B.}~\bibnamefont {Büchner}}, \bibinfo {author}
  {\bibfnamefont {M.~H.}\ \bibnamefont {Rümmeli}}, \ and\ \bibinfo {author}
  {\bibfnamefont {J.~H.}\ \bibnamefont {Warner}},\ }\href {\doibase
  10.1021/nn202051g} {\bibfield  {journal} {\bibinfo  {journal} {ACS Nano}\
  }\textbf {\bibinfo {volume} {5}},\ \bibinfo {pages} {6610} (\bibinfo {year}
  {2011})}\BibitemShut {NoStop}%
\bibitem [{\citenamefont {Britnell}\ \emph {et~al.}(2012)\citenamefont
  {Britnell}, \citenamefont {Gorbachev}, \citenamefont {Jalil}, \citenamefont
  {Belle}, \citenamefont {Schedin}, \citenamefont {Mishchenko}, \citenamefont
  {Georgiou}, \citenamefont {Katsnelson}, \citenamefont {Eaves}, \citenamefont
  {Morozov} \emph {et~al.}}]{junction1}%
  \BibitemOpen
  \bibfield  {author} {\bibinfo {author} {\bibfnamefont {L.}~\bibnamefont
  {Britnell}}, \bibinfo {author} {\bibfnamefont {R.}~\bibnamefont {Gorbachev}},
  \bibinfo {author} {\bibfnamefont {R.}~\bibnamefont {Jalil}}, \bibinfo
  {author} {\bibfnamefont {B.}~\bibnamefont {Belle}}, \bibinfo {author}
  {\bibfnamefont {F.}~\bibnamefont {Schedin}}, \bibinfo {author} {\bibfnamefont
  {A.}~\bibnamefont {Mishchenko}}, \bibinfo {author} {\bibfnamefont
  {T.}~\bibnamefont {Georgiou}}, \bibinfo {author} {\bibfnamefont
  {M.}~\bibnamefont {Katsnelson}}, \bibinfo {author} {\bibfnamefont
  {L.}~\bibnamefont {Eaves}}, \bibinfo {author} {\bibfnamefont
  {S.}~\bibnamefont {Morozov}},  \emph {et~al.},\ }\href@noop {} {\bibfield
  {journal} {\bibinfo  {journal} {Science}\ }\textbf {\bibinfo {volume}
  {335}},\ \bibinfo {pages} {947} (\bibinfo {year} {2012})}\BibitemShut
  {NoStop}%
\bibitem [{\citenamefont {Seo}\ \emph {et~al.}(2013)\citenamefont {Seo},
  \citenamefont {Min}, \citenamefont {Lee},\ and\ \citenamefont
  {Lee}}]{junction2}%
  \BibitemOpen
  \bibfield  {author} {\bibinfo {author} {\bibfnamefont {S.}~\bibnamefont
  {Seo}}, \bibinfo {author} {\bibfnamefont {M.}~\bibnamefont {Min}}, \bibinfo
  {author} {\bibfnamefont {S.~M.}\ \bibnamefont {Lee}}, \ and\ \bibinfo
  {author} {\bibfnamefont {H.}~\bibnamefont {Lee}},\ }\href@noop {} {\bibfield
  {journal} {\bibinfo  {journal} {Nature communications}\ }\textbf {\bibinfo
  {volume} {4}},\ \bibinfo {pages} {1920} (\bibinfo {year} {2013})}\BibitemShut
  {NoStop}%
\bibitem [{\citenamefont {Landauer}(1957)}]{laudau}%
  \BibitemOpen
  \bibfield  {author} {\bibinfo {author} {\bibfnamefont {R.}~\bibnamefont
  {Landauer}},\ }\href {\doibase 10.1147/rd.13.0223} {\bibfield  {journal}
  {\bibinfo  {journal} {IBM Journal of Research and Development}\ }\textbf
  {\bibinfo {volume} {1}},\ \bibinfo {pages} {223} (\bibinfo {year}
  {1957})}\BibitemShut {NoStop}%
\bibitem [{\citenamefont {Buttiker}(1988)}]{buttiker}%
  \BibitemOpen
  \bibfield  {author} {\bibinfo {author} {\bibfnamefont {M.}~\bibnamefont
  {Buttiker}},\ }\href {\doibase 10.1147/rd.323.0317} {\bibfield  {journal}
  {\bibinfo  {journal} {IBM Journal of Research and Development}\ }\textbf
  {\bibinfo {volume} {32}},\ \bibinfo {pages} {317} (\bibinfo {year}
  {1988})}\BibitemShut {NoStop}%
\bibitem [{\citenamefont {Mujica}\ \emph {et~al.}(1994)\citenamefont {Mujica},
  \citenamefont {Kemp},\ and\ \citenamefont {Ratner}}]{local1}%
  \BibitemOpen
  \bibfield  {author} {\bibinfo {author} {\bibfnamefont {V.}~\bibnamefont
  {Mujica}}, \bibinfo {author} {\bibfnamefont {M.}~\bibnamefont {Kemp}}, \ and\
  \bibinfo {author} {\bibfnamefont {M.~A.}\ \bibnamefont {Ratner}},\ }\href
  {\doibase http://dx.doi.org/10.1063/1.468314} {\bibfield  {journal} {\bibinfo
   {journal} {The Journal of Chemical Physics}\ }\textbf {\bibinfo {volume}
  {101}},\ \bibinfo {pages} {6849} (\bibinfo {year} {1994})}\BibitemShut
  {NoStop}%
\bibitem [{\citenamefont {Taylor}\ \emph {et~al.}(2001)\citenamefont {Taylor},
  \citenamefont {Guo},\ and\ \citenamefont {Wang}}]{local2}%
  \BibitemOpen
  \bibfield  {author} {\bibinfo {author} {\bibfnamefont {J.}~\bibnamefont
  {Taylor}}, \bibinfo {author} {\bibfnamefont {H.}~\bibnamefont {Guo}}, \ and\
  \bibinfo {author} {\bibfnamefont {J.}~\bibnamefont {Wang}},\ }\href {\doibase
  10.1103/PhysRevB.63.245407} {\bibfield  {journal} {\bibinfo  {journal} {Phys.
  Rev. B}\ }\textbf {\bibinfo {volume} {63}},\ \bibinfo {pages} {245407}
  (\bibinfo {year} {2001})}\BibitemShut {NoStop}%
\bibitem [{\citenamefont {Brandbyge}\ \emph {et~al.}(2002)\citenamefont
  {Brandbyge}, \citenamefont {Mozos}, \citenamefont {Ordej\'on}, \citenamefont
  {Taylor},\ and\ \citenamefont {Stokbro}}]{local3}%
  \BibitemOpen
  \bibfield  {author} {\bibinfo {author} {\bibfnamefont {M.}~\bibnamefont
  {Brandbyge}}, \bibinfo {author} {\bibfnamefont {J.-L.}\ \bibnamefont
  {Mozos}}, \bibinfo {author} {\bibfnamefont {P.}~\bibnamefont {Ordej\'on}},
  \bibinfo {author} {\bibfnamefont {J.}~\bibnamefont {Taylor}}, \ and\ \bibinfo
  {author} {\bibfnamefont {K.}~\bibnamefont {Stokbro}},\ }\href {\doibase
  10.1103/PhysRevB.65.165401} {\bibfield  {journal} {\bibinfo  {journal} {Phys.
  Rev. B}\ }\textbf {\bibinfo {volume} {65}},\ \bibinfo {pages} {165401}
  (\bibinfo {year} {2002})}\BibitemShut {NoStop}%
\bibitem [{\citenamefont {Xue}\ \emph {et~al.}(2002)\citenamefont {Xue},
  \citenamefont {Datta},\ and\ \citenamefont {Ratner}}]{local4}%
  \BibitemOpen
  \bibfield  {author} {\bibinfo {author} {\bibfnamefont {Y.}~\bibnamefont
  {Xue}}, \bibinfo {author} {\bibfnamefont {S.}~\bibnamefont {Datta}}, \ and\
  \bibinfo {author} {\bibfnamefont {M.~A.}\ \bibnamefont {Ratner}},\ }\href
  {\doibase http://dx.doi.org/10.1016/S0301-0104(02)00446-9} {\bibfield
  {journal} {\bibinfo  {journal} {Chemical Physics}\ }\textbf {\bibinfo
  {volume} {281}},\ \bibinfo {pages} {151 } (\bibinfo {year}
  {2002})}\BibitemShut {NoStop}%
\bibitem [{\citenamefont {Zhang}\ \emph {et~al.}(2004)\citenamefont {Zhang},
  \citenamefont {Zhang}, \citenamefont {Krsti\ifmmode~\acute{c}\else
  \'{c}\fi{}}, \citenamefont {Cheng}, \citenamefont {Butler},\ and\
  \citenamefont {MacLaren}}]{green-scatter}%
  \BibitemOpen
  \bibfield  {author} {\bibinfo {author} {\bibfnamefont {C.}~\bibnamefont
  {Zhang}}, \bibinfo {author} {\bibfnamefont {X.-G.}\ \bibnamefont {Zhang}},
  \bibinfo {author} {\bibfnamefont {P.~S.}\ \bibnamefont
  {Krsti\ifmmode~\acute{c}\else \'{c}\fi{}}}, \bibinfo {author} {\bibfnamefont
  {H.-p.}\ \bibnamefont {Cheng}}, \bibinfo {author} {\bibfnamefont {W.~H.}\
  \bibnamefont {Butler}}, \ and\ \bibinfo {author} {\bibfnamefont {J.~M.}\
  \bibnamefont {MacLaren}},\ }\href {\doibase 10.1103/PhysRevB.69.134406}
  {\bibfield  {journal} {\bibinfo  {journal} {Phys. Rev. B}\ }\textbf {\bibinfo
  {volume} {69}},\ \bibinfo {pages} {134406} (\bibinfo {year}
  {2004})}\BibitemShut {NoStop}%
\bibitem [{\citenamefont {Joon~Choi}\ and\ \citenamefont
  {Ihm}(1999)}]{pwcond1}%
  \BibitemOpen
  \bibfield  {author} {\bibinfo {author} {\bibfnamefont {H.}~\bibnamefont
  {Joon~Choi}}\ and\ \bibinfo {author} {\bibfnamefont {J.}~\bibnamefont
  {Ihm}},\ }\href {\doibase 10.1103/PhysRevB.59.2267} {\bibfield  {journal}
  {\bibinfo  {journal} {Phys. Rev. B}\ }\textbf {\bibinfo {volume} {59}},\
  \bibinfo {pages} {2267} (\bibinfo {year} {1999})}\BibitemShut {NoStop}%
\bibitem [{\citenamefont {Choi}\ \emph {et~al.}(2000)\citenamefont {Choi},
  \citenamefont {Ihm}, \citenamefont {Louie},\ and\ \citenamefont
  {Cohen}}]{pwcond2}%
  \BibitemOpen
  \bibfield  {author} {\bibinfo {author} {\bibfnamefont {H.~J.}\ \bibnamefont
  {Choi}}, \bibinfo {author} {\bibfnamefont {J.}~\bibnamefont {Ihm}}, \bibinfo
  {author} {\bibfnamefont {S.~G.}\ \bibnamefont {Louie}}, \ and\ \bibinfo
  {author} {\bibfnamefont {M.~L.}\ \bibnamefont {Cohen}},\ }\href {\doibase
  10.1103/PhysRevLett.84.2917} {\bibfield  {journal} {\bibinfo  {journal}
  {Phys. Rev. Lett.}\ }\textbf {\bibinfo {volume} {84}},\ \bibinfo {pages}
  {2917} (\bibinfo {year} {2000})}\BibitemShut {NoStop}%
\bibitem [{\citenamefont {Giannozzi}\ \emph {et~al.}(2009)\citenamefont
  {Giannozzi}, \citenamefont {Baroni}, \citenamefont {Bonini}, \citenamefont
  {Calandra}, \citenamefont {Car}, \citenamefont {Cavazzoni}, \citenamefont
  {Ceresoli}, \citenamefont {Chiarotti}, \citenamefont {Cococcioni},
  \citenamefont {Dabo}, \citenamefont {Corso}, \citenamefont {de~Gironcoli},
  \citenamefont {Fabris}, \citenamefont {Fratesi}, \citenamefont {Gebauer},
  \citenamefont {Gerstmann}, \citenamefont {Gougoussis}, \citenamefont
  {Kokalj}, \citenamefont {Lazzeri}, \citenamefont {Martin-Samos},
  \citenamefont {Marzari}, \citenamefont {Mauri}, \citenamefont {Mazzarello},
  \citenamefont {Paolini}, \citenamefont {Pasquarello}, \citenamefont
  {Paulatto}, \citenamefont {Sbraccia}, \citenamefont {Scandolo}, \citenamefont
  {Sclauzero}, \citenamefont {Seitsonen}, \citenamefont {Smogunov},
  \citenamefont {Umari},\ and\ \citenamefont {Wentzcovitch}}]{QM}%
  \BibitemOpen
  \bibfield  {author} {\bibinfo {author} {\bibfnamefont {P.}~\bibnamefont
  {Giannozzi}}, \bibinfo {author} {\bibfnamefont {S.}~\bibnamefont {Baroni}},
  \bibinfo {author} {\bibfnamefont {N.}~\bibnamefont {Bonini}}, \bibinfo
  {author} {\bibfnamefont {M.}~\bibnamefont {Calandra}}, \bibinfo {author}
  {\bibfnamefont {R.}~\bibnamefont {Car}}, \bibinfo {author} {\bibfnamefont
  {C.}~\bibnamefont {Cavazzoni}}, \bibinfo {author} {\bibfnamefont
  {D.}~\bibnamefont {Ceresoli}}, \bibinfo {author} {\bibfnamefont {G.~L.}\
  \bibnamefont {Chiarotti}}, \bibinfo {author} {\bibfnamefont {M.}~\bibnamefont
  {Cococcioni}}, \bibinfo {author} {\bibfnamefont {I.}~\bibnamefont {Dabo}},
  \bibinfo {author} {\bibfnamefont {A.~D.}\ \bibnamefont {Corso}}, \bibinfo
  {author} {\bibfnamefont {S.}~\bibnamefont {de~Gironcoli}}, \bibinfo {author}
  {\bibfnamefont {S.}~\bibnamefont {Fabris}}, \bibinfo {author} {\bibfnamefont
  {G.}~\bibnamefont {Fratesi}}, \bibinfo {author} {\bibfnamefont
  {R.}~\bibnamefont {Gebauer}}, \bibinfo {author} {\bibfnamefont
  {U.}~\bibnamefont {Gerstmann}}, \bibinfo {author} {\bibfnamefont
  {C.}~\bibnamefont {Gougoussis}}, \bibinfo {author} {\bibfnamefont
  {A.}~\bibnamefont {Kokalj}}, \bibinfo {author} {\bibfnamefont
  {M.}~\bibnamefont {Lazzeri}}, \bibinfo {author} {\bibfnamefont
  {L.}~\bibnamefont {Martin-Samos}}, \bibinfo {author} {\bibfnamefont
  {N.}~\bibnamefont {Marzari}}, \bibinfo {author} {\bibfnamefont
  {F.}~\bibnamefont {Mauri}}, \bibinfo {author} {\bibfnamefont
  {R.}~\bibnamefont {Mazzarello}}, \bibinfo {author} {\bibfnamefont
  {S.}~\bibnamefont {Paolini}}, \bibinfo {author} {\bibfnamefont
  {A.}~\bibnamefont {Pasquarello}}, \bibinfo {author} {\bibfnamefont
  {L.}~\bibnamefont {Paulatto}}, \bibinfo {author} {\bibfnamefont
  {C.}~\bibnamefont {Sbraccia}}, \bibinfo {author} {\bibfnamefont
  {S.}~\bibnamefont {Scandolo}}, \bibinfo {author} {\bibfnamefont
  {G.}~\bibnamefont {Sclauzero}}, \bibinfo {author} {\bibfnamefont {A.~P.}\
  \bibnamefont {Seitsonen}}, \bibinfo {author} {\bibfnamefont {A.}~\bibnamefont
  {Smogunov}}, \bibinfo {author} {\bibfnamefont {P.}~\bibnamefont {Umari}}, \
  and\ \bibinfo {author} {\bibfnamefont {R.~M.}\ \bibnamefont {Wentzcovitch}},\
  }\href {http://stacks.iop.org/0953-8984/21/i=39/a=395502} {\bibfield
  {journal} {\bibinfo  {journal} {Journal of Physics: Condensed Matter}\
  }\textbf {\bibinfo {volume} {21}},\ \bibinfo {pages} {395502} (\bibinfo
  {year} {2009})}\BibitemShut {NoStop}%
\bibitem [{\citenamefont {Smogunov}\ \emph
  {et~al.}(2004{\natexlab{a}})\citenamefont {Smogunov}, \citenamefont
  {Dal~Corso},\ and\ \citenamefont {Tosatti}}]{pwcondap1}%
  \BibitemOpen
  \bibfield  {author} {\bibinfo {author} {\bibfnamefont {A.}~\bibnamefont
  {Smogunov}}, \bibinfo {author} {\bibfnamefont {A.}~\bibnamefont {Dal~Corso}},
  \ and\ \bibinfo {author} {\bibfnamefont {E.}~\bibnamefont {Tosatti}},\ }\href
  {\doibase 10.1103/PhysRevB.70.045417} {\bibfield  {journal} {\bibinfo
  {journal} {Phys. Rev. B}\ }\textbf {\bibinfo {volume} {70}},\ \bibinfo
  {pages} {045417} (\bibinfo {year} {2004}{\natexlab{a}})}\BibitemShut
  {NoStop}%
\bibitem [{\citenamefont {Kuroda}\ \emph {et~al.}(2011)\citenamefont {Kuroda},
  \citenamefont {Tersoff}, \citenamefont {Newns},\ and\ \citenamefont
  {Martyna}}]{pwcondap5}%
  \BibitemOpen
  \bibfield  {author} {\bibinfo {author} {\bibfnamefont {M.~A.}\ \bibnamefont
  {Kuroda}}, \bibinfo {author} {\bibfnamefont {J.}~\bibnamefont {Tersoff}},
  \bibinfo {author} {\bibfnamefont {D.~M.}\ \bibnamefont {Newns}}, \ and\
  \bibinfo {author} {\bibfnamefont {G.~J.}\ \bibnamefont {Martyna}},\ }\href
  {\doibase 10.1021/nl201436b} {\bibfield  {journal} {\bibinfo  {journal} {Nano
  Letters}\ }\textbf {\bibinfo {volume} {11}},\ \bibinfo {pages} {3629}
  (\bibinfo {year} {2011})}\BibitemShut {NoStop}%
\bibitem [{\citenamefont {Yazyev}\ and\ \citenamefont
  {Pasquarello}(2009)}]{pwcondap3}%
  \BibitemOpen
  \bibfield  {author} {\bibinfo {author} {\bibfnamefont {O.~V.}\ \bibnamefont
  {Yazyev}}\ and\ \bibinfo {author} {\bibfnamefont {A.}~\bibnamefont
  {Pasquarello}},\ }\href {\doibase 10.1103/PhysRevB.80.035408} {\bibfield
  {journal} {\bibinfo  {journal} {Phys. Rev. B}\ }\textbf {\bibinfo {volume}
  {80}},\ \bibinfo {pages} {035408} (\bibinfo {year} {2009})}\BibitemShut
  {NoStop}%
\bibitem [{\citenamefont {Velev}\ \emph {et~al.}(2007)\citenamefont {Velev},
  \citenamefont {Duan}, \citenamefont {Belashchenko}, \citenamefont {Jaswal},\
  and\ \citenamefont {Tsymbal}}]{pwcondap4}%
  \BibitemOpen
  \bibfield  {author} {\bibinfo {author} {\bibfnamefont {J.~P.}\ \bibnamefont
  {Velev}}, \bibinfo {author} {\bibfnamefont {C.-G.}\ \bibnamefont {Duan}},
  \bibinfo {author} {\bibfnamefont {K.~D.}\ \bibnamefont {Belashchenko}},
  \bibinfo {author} {\bibfnamefont {S.~S.}\ \bibnamefont {Jaswal}}, \ and\
  \bibinfo {author} {\bibfnamefont {E.~Y.}\ \bibnamefont {Tsymbal}},\ }\href
  {\doibase 10.1103/PhysRevLett.98.137201} {\bibfield  {journal} {\bibinfo
  {journal} {Phys. Rev. Lett.}\ }\textbf {\bibinfo {volume} {98}},\ \bibinfo
  {pages} {137201} (\bibinfo {year} {2007})}\BibitemShut {NoStop}%
\bibitem [{\citenamefont {Sclauzero}\ \emph {et~al.}(2012)\citenamefont
  {Sclauzero}, \citenamefont {Dal~Corso},\ and\ \citenamefont
  {Smogunov}}]{pwcondap6}%
  \BibitemOpen
  \bibfield  {author} {\bibinfo {author} {\bibfnamefont {G.}~\bibnamefont
  {Sclauzero}}, \bibinfo {author} {\bibfnamefont {A.}~\bibnamefont
  {Dal~Corso}}, \ and\ \bibinfo {author} {\bibfnamefont {A.}~\bibnamefont
  {Smogunov}},\ }\href {\doibase 10.1103/PhysRevB.85.165412} {\bibfield
  {journal} {\bibinfo  {journal} {Phys. Rev. B}\ }\textbf {\bibinfo {volume}
  {85}},\ \bibinfo {pages} {165412} (\bibinfo {year} {2012})}\BibitemShut
  {NoStop}%
\bibitem [{\citenamefont {MacLaren}\ \emph {et~al.}(1999)\citenamefont
  {MacLaren}, \citenamefont {Zhang}, \citenamefont {Butler},\ and\
  \citenamefont {Wang}}]{lkkrap2}%
  \BibitemOpen
  \bibfield  {author} {\bibinfo {author} {\bibfnamefont {J.~M.}\ \bibnamefont
  {MacLaren}}, \bibinfo {author} {\bibfnamefont {X.-G.}\ \bibnamefont {Zhang}},
  \bibinfo {author} {\bibfnamefont {W.~H.}\ \bibnamefont {Butler}}, \ and\
  \bibinfo {author} {\bibfnamefont {X.}~\bibnamefont {Wang}},\ }\href {\doibase
  10.1103/PhysRevB.59.5470} {\bibfield  {journal} {\bibinfo  {journal} {Phys.
  Rev. B}\ }\textbf {\bibinfo {volume} {59}},\ \bibinfo {pages} {5470}
  (\bibinfo {year} {1999})}\BibitemShut {NoStop}%
\bibitem [{\citenamefont {Kim}\ \emph {et~al.}(2010)\citenamefont {Kim},
  \citenamefont {Zhang}, \citenamefont {Nicholson}, \citenamefont {Evans},
  \citenamefont {Kulkarni}, \citenamefont {Radhakrishnan}, \citenamefont
  {Kenik},\ and\ \citenamefont {Li}}]{lkkrapp1}%
  \BibitemOpen
  \bibfield  {author} {\bibinfo {author} {\bibfnamefont {T.-H.}\ \bibnamefont
  {Kim}}, \bibinfo {author} {\bibfnamefont {X.-G.}\ \bibnamefont {Zhang}},
  \bibinfo {author} {\bibfnamefont {D.~M.}\ \bibnamefont {Nicholson}}, \bibinfo
  {author} {\bibfnamefont {B.~M.}\ \bibnamefont {Evans}}, \bibinfo {author}
  {\bibfnamefont {N.~S.}\ \bibnamefont {Kulkarni}}, \bibinfo {author}
  {\bibfnamefont {B.}~\bibnamefont {Radhakrishnan}}, \bibinfo {author}
  {\bibfnamefont {E.~A.}\ \bibnamefont {Kenik}}, \ and\ \bibinfo {author}
  {\bibfnamefont {A.-P.}\ \bibnamefont {Li}},\ }\href@noop {} {\bibfield
  {journal} {\bibinfo  {journal} {Nano letters}\ }\textbf {\bibinfo {volume}
  {10}},\ \bibinfo {pages} {3096} (\bibinfo {year} {2010})}\BibitemShut
  {NoStop}%
\bibitem [{\citenamefont {Zhang}\ \emph {et~al.}(2010)\citenamefont {Zhang},
  \citenamefont {Wang}, \citenamefont {Zhang},\ and\ \citenamefont
  {Han}}]{lkkrapp2}%
  \BibitemOpen
  \bibfield  {author} {\bibinfo {author} {\bibfnamefont {J.}~\bibnamefont
  {Zhang}}, \bibinfo {author} {\bibfnamefont {Y.}~\bibnamefont {Wang}},
  \bibinfo {author} {\bibfnamefont {X.-G.}\ \bibnamefont {Zhang}}, \ and\
  \bibinfo {author} {\bibfnamefont {X.~F.}\ \bibnamefont {Han}},\ }\href
  {\doibase 10.1103/PhysRevB.82.134449} {\bibfield  {journal} {\bibinfo
  {journal} {Phys. Rev. B}\ }\textbf {\bibinfo {volume} {82}},\ \bibinfo
  {pages} {134449} (\bibinfo {year} {2010})}\BibitemShut {NoStop}%
\bibitem [{\citenamefont {Wang}\ \emph {et~al.}(2010)\citenamefont {Wang},
  \citenamefont {Zhang}, \citenamefont {Zhang}, \citenamefont {Cheng},\ and\
  \citenamefont {Han}}]{lkkrapp3}%
  \BibitemOpen
  \bibfield  {author} {\bibinfo {author} {\bibfnamefont {Y.}~\bibnamefont
  {Wang}}, \bibinfo {author} {\bibfnamefont {J.}~\bibnamefont {Zhang}},
  \bibinfo {author} {\bibfnamefont {X.-G.}\ \bibnamefont {Zhang}}, \bibinfo
  {author} {\bibfnamefont {H.-P.}\ \bibnamefont {Cheng}}, \ and\ \bibinfo
  {author} {\bibfnamefont {X.~F.}\ \bibnamefont {Han}},\ }\href {\doibase
  10.1103/PhysRevB.82.054405} {\bibfield  {journal} {\bibinfo  {journal} {Phys.
  Rev. B}\ }\textbf {\bibinfo {volume} {82}},\ \bibinfo {pages} {054405}
  (\bibinfo {year} {2010})}\BibitemShut {NoStop}%
\bibitem [{\citenamefont {Zhang}\ \emph
  {et~al.}(2007{\natexlab{a}})\citenamefont {Zhang}, \citenamefont {Varga},\
  and\ \citenamefont {Pantelides}}]{cmband}%
  \BibitemOpen
  \bibfield  {author} {\bibinfo {author} {\bibfnamefont {X.-G.}\ \bibnamefont
  {Zhang}}, \bibinfo {author} {\bibfnamefont {K.}~\bibnamefont {Varga}}, \ and\
  \bibinfo {author} {\bibfnamefont {S.~T.}\ \bibnamefont {Pantelides}},\ }\href
  {\doibase 10.1103/PhysRevB.76.035108} {\bibfield  {journal} {\bibinfo
  {journal} {Phys. Rev. B}\ }\textbf {\bibinfo {volume} {76}},\ \bibinfo
  {pages} {035108} (\bibinfo {year} {2007}{\natexlab{a}})}\BibitemShut
  {NoStop}%
\bibitem [{\citenamefont {Srivastava}\ \emph {et~al.}(2012)\citenamefont
  {Srivastava}, \citenamefont {Wang}, \citenamefont {Zhang}, \citenamefont
  {Nicholson},\ and\ \citenamefont {Cheng}}]{Manoj}%
  \BibitemOpen
  \bibfield  {author} {\bibinfo {author} {\bibfnamefont {M.~K.}\ \bibnamefont
  {Srivastava}}, \bibinfo {author} {\bibfnamefont {Y.}~\bibnamefont {Wang}},
  \bibinfo {author} {\bibfnamefont {X.-G.}\ \bibnamefont {Zhang}}, \bibinfo
  {author} {\bibfnamefont {D.~M.~C.}\ \bibnamefont {Nicholson}}, \ and\
  \bibinfo {author} {\bibfnamefont {H.-P.}\ \bibnamefont {Cheng}},\ }\href
  {\doibase 10.1103/PhysRevB.86.075134} {\bibfield  {journal} {\bibinfo
  {journal} {Phys. Rev. B}\ }\textbf {\bibinfo {volume} {86}},\ \bibinfo
  {pages} {075134} (\bibinfo {year} {2012})}\BibitemShut {NoStop}%
\bibitem [{\citenamefont {Kohn}\ and\ \citenamefont {Sham}(1965)}]{DFT}%
  \BibitemOpen
  \bibfield  {author} {\bibinfo {author} {\bibfnamefont {W.}~\bibnamefont
  {Kohn}}\ and\ \bibinfo {author} {\bibfnamefont {L.~J.}\ \bibnamefont
  {Sham}},\ }\href {\doibase 10.1103/PhysRev.140.A1133} {\bibfield  {journal}
  {\bibinfo  {journal} {Phys. Rev.}\ }\textbf {\bibinfo {volume} {140}},\
  \bibinfo {pages} {A1133} (\bibinfo {year} {1965})}\BibitemShut {NoStop}%
\bibitem [{\citenamefont {Smogunov}\ \emph
  {et~al.}(2004{\natexlab{b}})\citenamefont {Smogunov}, \citenamefont
  {Dal~Corso},\ and\ \citenamefont {Tosatti}}]{altrasoft}%
  \BibitemOpen
  \bibfield  {author} {\bibinfo {author} {\bibfnamefont {A.}~\bibnamefont
  {Smogunov}}, \bibinfo {author} {\bibfnamefont {A.}~\bibnamefont {Dal~Corso}},
  \ and\ \bibinfo {author} {\bibfnamefont {E.}~\bibnamefont {Tosatti}},\ }\href
  {\doibase 10.1103/PhysRevB.70.045417} {\bibfield  {journal} {\bibinfo
  {journal} {Phys. Rev. B}\ }\textbf {\bibinfo {volume} {70}},\ \bibinfo
  {pages} {045417} (\bibinfo {year} {2004}{\natexlab{b}})}\BibitemShut
  {NoStop}%
\bibitem [{\citenamefont {Troullier}\ and\ \citenamefont
  {Martins}(1991)}]{NCPP}%
  \BibitemOpen
  \bibfield  {author} {\bibinfo {author} {\bibfnamefont {N.}~\bibnamefont
  {Troullier}}\ and\ \bibinfo {author} {\bibfnamefont {J.~L.}\ \bibnamefont
  {Martins}},\ }\href@noop {} {\bibfield  {journal} {\bibinfo  {journal} {Phys.
  Rev. B}\ }\textbf {\bibinfo {volume} {43}},\ \bibinfo {pages} {1993}
  (\bibinfo {year} {1991})}\BibitemShut {NoStop}%
\bibitem [{\citenamefont {Born}\ and\ \citenamefont {Wolf}(1964)}]{tranM}%
  \BibitemOpen
  \bibfield  {author} {\bibinfo {author} {\bibfnamefont {M.}~\bibnamefont
  {Born}}\ and\ \bibinfo {author} {\bibfnamefont {E.}~\bibnamefont {Wolf}},\
  }\href@noop {} {\emph {\bibinfo {title} {Principles of optics:
  electromagnetic theory of propagation, interference and diffraction of
  light}}}\ (\bibinfo  {publisher} {Oxford, Pergamon Press},\ \bibinfo {year}
  {1964})\BibitemShut {NoStop}%
\bibitem [{\citenamefont {Ko}\ and\ \citenamefont {Sambles}(1988)}]{num}%
  \BibitemOpen
  \bibfield  {author} {\bibinfo {author} {\bibfnamefont {D.~Y.~K.}\
  \bibnamefont {Ko}}\ and\ \bibinfo {author} {\bibfnamefont {J.~R.}\
  \bibnamefont {Sambles}},\ }\href {\doibase 10.1364/JOSAA.5.001863} {\bibfield
   {journal} {\bibinfo  {journal} {J. Opt. Soc. Am. A}\ }\textbf {\bibinfo
  {volume} {5}},\ \bibinfo {pages} {1863} (\bibinfo {year} {1988})}\BibitemShut
  {NoStop}%
\bibitem [{\citenamefont {Cotter}\ \emph {et~al.}(1995)\citenamefont {Cotter},
  \citenamefont {Preist},\ and\ \citenamefont {Sambles}}]{scat}%
  \BibitemOpen
  \bibfield  {author} {\bibinfo {author} {\bibfnamefont {N.~P.~K.}\
  \bibnamefont {Cotter}}, \bibinfo {author} {\bibfnamefont {T.~W.}\
  \bibnamefont {Preist}}, \ and\ \bibinfo {author} {\bibfnamefont {J.~R.}\
  \bibnamefont {Sambles}},\ }\href {\doibase 10.1364/JOSAA.12.001097}
  {\bibfield  {journal} {\bibinfo  {journal} {J. Opt. Soc. Am. A}\ }\textbf
  {\bibinfo {volume} {12}},\ \bibinfo {pages} {1097} (\bibinfo {year}
  {1995})}\BibitemShut {NoStop}%
\bibitem [{\citenamefont {Wheeler}(1937)}]{scat1}%
  \BibitemOpen
  \bibfield  {author} {\bibinfo {author} {\bibfnamefont {J.~A.}\ \bibnamefont
  {Wheeler}},\ }\href {\doibase 10.1103/PhysRev.52.1107} {\bibfield  {journal}
  {\bibinfo  {journal} {Phys. Rev.}\ }\textbf {\bibinfo {volume} {52}},\
  \bibinfo {pages} {1107} (\bibinfo {year} {1937})}\BibitemShut {NoStop}%
\bibitem [{\citenamefont {Merzbacher}(1970)}]{scat2}%
  \BibitemOpen
  \bibfield  {author} {\bibinfo {author} {\bibfnamefont {E.}~\bibnamefont
  {Merzbacher}},\ }\href@noop {} {\emph {\bibinfo {title} {Quantum Mechanics,
  Chap.6, Sec. 6}}}\ (\bibinfo  {publisher} {Wiley, New York},\ \bibinfo {year}
  {1970})\BibitemShut {NoStop}%
\bibitem [{\citenamefont {MacLaren}\ \emph {et~al.}(1990)\citenamefont
  {MacLaren}, \citenamefont {Crampin}, \citenamefont {Vvedensky}, \citenamefont
  {Albers},\ and\ \citenamefont {Pendry}}]{lkk}%
  \BibitemOpen
  \bibfield  {author} {\bibinfo {author} {\bibfnamefont {J.}~\bibnamefont
  {MacLaren}}, \bibinfo {author} {\bibfnamefont {S.}~\bibnamefont {Crampin}},
  \bibinfo {author} {\bibfnamefont {D.}~\bibnamefont {Vvedensky}}, \bibinfo
  {author} {\bibfnamefont {R.}~\bibnamefont {Albers}}, \ and\ \bibinfo {author}
  {\bibfnamefont {J.}~\bibnamefont {Pendry}},\ }\href {\doibase
  http://dx.doi.org/10.1016/0010-4655(90)90035-Y} {\bibfield  {journal}
  {\bibinfo  {journal} {Computer Physics Communications}\ }\textbf {\bibinfo
  {volume} {60}},\ \bibinfo {pages} {365 } (\bibinfo {year}
  {1990})}\BibitemShut {NoStop}%
\bibitem [{\citenamefont {Zhang}\ \emph
  {et~al.}(2007{\natexlab{b}})\citenamefont {Zhang}, \citenamefont {Varga},\
  and\ \citenamefont {Pantelides}}]{bd-zhang}%
  \BibitemOpen
  \bibfield  {author} {\bibinfo {author} {\bibfnamefont {X.-G.}\ \bibnamefont
  {Zhang}}, \bibinfo {author} {\bibfnamefont {K.}~\bibnamefont {Varga}}, \ and\
  \bibinfo {author} {\bibfnamefont {S.~T.}\ \bibnamefont {Pantelides}},\ }\href
  {\doibase 10.1103/PhysRevB.76.035108} {\bibfield  {journal} {\bibinfo
  {journal} {Phys. Rev. B}\ }\textbf {\bibinfo {volume} {76}},\ \bibinfo
  {pages} {035108} (\bibinfo {year} {2007}{\natexlab{b}})}\BibitemShut
  {NoStop}%
\bibitem [{\citenamefont {Rappe}\ \emph {et~al.}(1990)\citenamefont {Rappe},
  \citenamefont {Rabe}, \citenamefont {Kaxiras},\ and\ \citenamefont
  {Joannopoulos}}]{rrkj}%
  \BibitemOpen
  \bibfield  {author} {\bibinfo {author} {\bibfnamefont {A.~M.}\ \bibnamefont
  {Rappe}}, \bibinfo {author} {\bibfnamefont {K.~M.}\ \bibnamefont {Rabe}},
  \bibinfo {author} {\bibfnamefont {E.}~\bibnamefont {Kaxiras}}, \ and\
  \bibinfo {author} {\bibfnamefont {J.~D.}\ \bibnamefont {Joannopoulos}},\
  }\href {\doibase 10.1103/PhysRevB.41.1227} {\bibfield  {journal} {\bibinfo
  {journal} {Phys. Rev. B}\ }\textbf {\bibinfo {volume} {41}},\ \bibinfo
  {pages} {1227} (\bibinfo {year} {1990})}\BibitemShut {NoStop}%
\bibitem [{\citenamefont {Perdew}\ \emph {et~al.}(1996)\citenamefont {Perdew},
  \citenamefont {Burke},\ and\ \citenamefont {Ernzerhof}}]{pbe}%
  \BibitemOpen
  \bibfield  {author} {\bibinfo {author} {\bibfnamefont {J.~P.}\ \bibnamefont
  {Perdew}}, \bibinfo {author} {\bibfnamefont {K.}~\bibnamefont {Burke}}, \
  and\ \bibinfo {author} {\bibfnamefont {M.}~\bibnamefont {Ernzerhof}},\ }\href
  {\doibase 10.1103/PhysRevLett.77.3865} {\bibfield  {journal} {\bibinfo
  {journal} {Phys. Rev. Lett.}\ }\textbf {\bibinfo {volume} {77}},\ \bibinfo
  {pages} {3865} (\bibinfo {year} {1996})}\BibitemShut {NoStop}%
\bibitem [{\citenamefont {Son}\ \emph {et~al.}(2006)\citenamefont {Son},
  \citenamefont {Cohen},\ and\ \citenamefont {Louie}}]{ZGNR}%
  \BibitemOpen
  \bibfield  {author} {\bibinfo {author} {\bibfnamefont {Y.-W.}\ \bibnamefont
  {Son}}, \bibinfo {author} {\bibfnamefont {M.~L.}\ \bibnamefont {Cohen}}, \
  and\ \bibinfo {author} {\bibfnamefont {S.~G.}\ \bibnamefont {Louie}},\ }\href
  {\doibase 10.1103/PhysRevLett.97.216803} {\bibfield  {journal} {\bibinfo
  {journal} {Phys. Rev. Lett.}\ }\textbf {\bibinfo {volume} {97}},\ \bibinfo
  {pages} {216803} (\bibinfo {year} {2006})}\BibitemShut {NoStop}%
\bibitem [{\citenamefont {Fisher}\ and\ \citenamefont {Lee}(1981)}]{spin2}%
  \BibitemOpen
  \bibfield  {author} {\bibinfo {author} {\bibfnamefont {D.~S.}\ \bibnamefont
  {Fisher}}\ and\ \bibinfo {author} {\bibfnamefont {P.~A.}\ \bibnamefont
  {Lee}},\ }\href@noop {} {\bibfield  {journal} {\bibinfo  {journal} {Phys.
  Rev. B}\ }\textbf {\bibinfo {volume} {23}},\ \bibinfo {pages} {6851}
  (\bibinfo {year} {1981})}\BibitemShut {NoStop}%
\bibitem [{\citenamefont {Cao}\ \emph {et~al.}(2011)\citenamefont {Cao},
  \citenamefont {Wang}, \citenamefont {Cheng},\ and\ \citenamefont
  {Jiang}}]{spin1}%
  \BibitemOpen
  \bibfield  {author} {\bibinfo {author} {\bibfnamefont {C.}~\bibnamefont
  {Cao}}, \bibinfo {author} {\bibfnamefont {Y.}~\bibnamefont {Wang}}, \bibinfo
  {author} {\bibfnamefont {H.-P.}\ \bibnamefont {Cheng}}, \ and\ \bibinfo
  {author} {\bibfnamefont {J.-Z.}\ \bibnamefont {Jiang}},\ }\href {\doibase
  http://dx.doi.org/10.1063/1.3626596} {\bibfield  {journal} {\bibinfo
  {journal} {Applied Physics Letters}\ }\textbf {\bibinfo {volume} {99}},\
  \bibinfo {pages} {073110} (\bibinfo {year} {2011})}\BibitemShut {NoStop}%
\bibitem [{\citenamefont {Trickey}\ \emph {et~al.}(1992)\citenamefont
  {Trickey}, \citenamefont {M\"uller-Plathe}, \citenamefont {Diercksen},\ and\
  \citenamefont {Boettger}}]{interlayer}%
  \BibitemOpen
  \bibfield  {author} {\bibinfo {author} {\bibfnamefont {S.~B.}\ \bibnamefont
  {Trickey}}, \bibinfo {author} {\bibfnamefont {F.}~\bibnamefont
  {M\"uller-Plathe}}, \bibinfo {author} {\bibfnamefont {G.~H.~F.}\ \bibnamefont
  {Diercksen}}, \ and\ \bibinfo {author} {\bibfnamefont {J.~C.}\ \bibnamefont
  {Boettger}},\ }\href {\doibase 10.1103/PhysRevB.45.4460} {\bibfield
  {journal} {\bibinfo  {journal} {Phys. Rev. B}\ }\textbf {\bibinfo {volume}
  {45}},\ \bibinfo {pages} {4460} (\bibinfo {year} {1992})}\BibitemShut
  {NoStop}%
\end{thebibliography}%
\end {document}